\documentclass{article}

\newif\ifpreprintmode

\usepackage[sglblindworkshop, final]{neurips_2025}
\workshoptitle{SimBioChem 2025}

\usepackage[pdftex]{graphicx}
\usepackage[utf8]{inputenc} 
\usepackage[T1]{fontenc}    
\usepackage[hidelinks]{hyperref}
\usepackage{url}            
\usepackage{booktabs}       
\usepackage{amsfonts}       
\usepackage{nicefrac}       
\usepackage{microtype}      
\usepackage{algorithm}
\usepackage{algpseudocode}
\usepackage{amsmath}
\usepackage{chemfig}
\usepackage{amssymb} 
\usepackage{multirow}
\usepackage{mhchem}
\usepackage{tablefootnote}
\usepackage{caption}

\usepackage{tikz}
\usetikzlibrary{arrows.meta, positioning, calc, decorations.pathreplacing}
\newcommand\markerlessfootnote[1]{%
  \begingroup
  \renewcommand\thefootnote{}\footnote{#1}%
  \addtocounter{footnote}{-1}%
  \endgroup
}

\title{Discovery of Sustainable Refrigerants through Physics-Informed RL Fine-Tuning of Sequence Models}

\author{%
    Adrien Goldszal$^{1,2,3\,}$\footnotemark[2]\,\,\,\footnotemark[1] \quad
    Diego Calanzone$^{2,3}$\footnotemark[1] \quad
    Vincent Taboga$^{2,3}$ \quad
    Pierre\mbox{-}Luc Bacon$^{2,3}$\\
    $^1$École Polytechnique \quad $^2$Mila Quebec AI Institute \quad $^3$Université de Montréal
}

\begin{document}
\maketitle

\ifpreprintmode
  \begingroup
    \renewcommand{\thefootnote}{\textdagger}%
    \footnotetext[2]{Work done during an internship at Mila.}%
  \endgroup
  \begingroup
    \renewcommand{\thefootnote}{\textasteriskcentered}%
    \footnotetext[1]{Corresponding authors: \texttt{adrien.goldszal@polytechnique.edu, diego.calanzone@mila.quebec}}%
  \endgroup
\fi


\begin{abstract}
Most refrigerants currently used in air-conditioning systems, such as hydrofluorocarbons, are potent greenhouse gases and are being phased down. Large-scale molecular screening has been applied to the search for alternatives, but in practice only about 300 refrigerants are known, and only a few additional candidates have been suggested without experimental validation. This scarcity of reliable data limits the effectiveness of purely data-driven methods. We present \texttt{Refgen}, a generative pipeline that integrates machine learning with physics-grounded inductive biases. Alongside fine-tuning for valid molecular generation, \texttt{Refgen}\markerlessfootnote{Code is available at \href{https://github.com/ddidacus/refgen}{https://github.com/ddidacus/refgen}.} incorporates predictive models for critical properties, equations of state, thermochemical polynomials, and full vapor compression cycle simulations. These models enable reinforcement learning fine-tuning under thermodynamic constraints, enforcing consistency and guiding discovery toward molecules that balance efficiency, safety, and environmental impact. By embedding physics into the learning process, \texttt{Refgen} leverages scarce data effectively and enables de novo refrigerant discovery beyond the known set of compounds.
\end{abstract}

\section{Introduction}

The search for new refrigerants involves balancing often competing objectives: achieving optimal thermodynamic properties, fluid behavior, and heat transfer efficiency while meeting environmental and safety requirements such as low Global Warming Potential (GWP), reduced flammability, and minimal toxicity. The landscape of refrigerant compounds has undergone major shifts in recent decades with the replacement of ozone depleting chlorofluorocarbons (CFC's) with hydrofluorocarbons (HFC's) through the Montréal Protocol \citep{UNEP_MontrealProtocol}; though, as many HFC's exhibit high GWP, recent efforts such as the Kigali Amendment \citep{kigali2016amendment} are organizing their phase down. Hydrofluoroolpehins (HFO's) are emerging as potential viable candidates, but long term environmental impact studies suggest that most heavily fluorinated compounds such as per- and polyfluoroalkyl substances (PFAS), which include many HFO's, should ideally be eliminated in favor of alternatives \citep{oecd2021reconciling, Gluge2024NonFluorinated}. While increasing interest is therefore taken in natural refrigerant options such as \ce{CO2} and propane, they present challenges of their own, with propane being highly flammable and \ce{CO2} operating at high pressures. 

Traditional \textit{high-throughput screening} has long been a staple in drug and materials discovery, where large molecular databases are filtered down to a small number of promising candidates through property prediction and expert knowledge before experimental validation. In the refrigerant domain, however, such systematic screening efforts have been rare. One of the few large-scale studies of this kind  \citep{McLinden2017} highlighted how limited the chemical space becomes once thermodynamic performance, safety, and environmental constraints are all enforced, identifying only a handful of plausible candidates. Moreover, accurate evaluation of molecular properties for this filtering process remains difficult due to the scarcity of publicly available ground-truth data. This motivates the recent shift toward learning predictive relationships between physicochemical properties and molecular structure as a way to support refrigerant discovery \citep{Kazakov2012ComputationalDO}.


In this work we explore a novel application of large molecule sequence models\footnote{A review of related work can be found in appendix \ref{appendix:related_work}}. We present \texttt{Refgen}, a framework for the discovery of new refrigerant candidates adapted to the challenges faced by the industry:
\begin{enumerate}
    \item We develop state-of-the-art physics-grounded property predictors trained independently from supervised datasets and physics models (Equations of state (EOS), NASA polynomials, group-contribution methods) to compute key thermodynamic and chemical properties of molecules. 
    \item We build a multi-reward RL pipeline using the property predictors to guide our LLM towards the generation of optimal refrigerant candidates in SMILES format.
\end{enumerate}

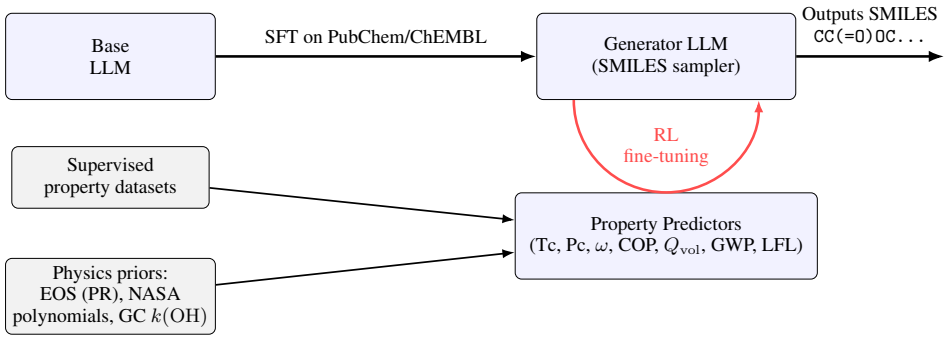
\begin{figure}[ht]
  \centering
  \resizebox{0.9\textwidth}{!}{%
\begin{tikzpicture}[font=\small]
  \tikzset{
    box/.style={
      draw, rounded corners=3pt, fill=blue!5,
      minimum width=3.4cm, minimum height=1.4cm, align=center
    },
    small/.style={
      draw, rounded corners=3pt, fill=gray!10,
      minimum width=3.2cm, minimum height=1.0cm, align=center
    },
    arrow/.style={-{Latex[length=2.5mm]}, very thick},
    feed/.style={-{Latex[length=2.0mm]}, thick},
    rewardarrow/.style={-{Latex[length=2mm]}, very thick, draw=red!70},
  }
  \node[box] (base) {Base\\LLM};
  \node[box, right=5.2cm of base, minimum width=4.2cm] (gen) {Generator LLM\\(SMILES sampler)};
  \draw[arrow] (base) -- node[midway, above, yshift=2pt]{\footnotesize SFT on PubChem/ChEMBL} (gen);
  \node[box, below=1.5cm of gen, minimum width=4.9cm] (pred)
       {Property Predictors\\(Tc, Pc, $\omega$, COP, $Q_{\mathrm{vol}}$, GWP, LFL)};
  
  \draw[rewardarrow] (gen.south) +(-1.5,0) arc (180:360:1.5)
    node[midway, above=8pt, text=red!70, font=\footnotesize, align=center]{RL\\fine-tuning};
  
  \node[small, below=0.75cm of base] (data) {Supervised\\property datasets};
  \node[small, below=0.8cm of data, minimum width=3.4cm] (physics)
       {Physics priors:\\ EOS (PR), NASA\\polynomials, GC $k(\mathrm{OH})$};
  \draw[feed] (data) -- (pred);
  \draw[feed] (physics) -- (pred);
  \draw[arrow] (gen.east) -- ++(2.4,0)
    node[midway, above, yshift=2pt, align=center]
    {Outputs SMILES\\ \texttt{CC(=O)OC\ldots}};
\end{tikzpicture}

  }
  \caption{The \texttt{Refgen} framework. The LLM is supervised on molecular corpora for valid SMILES generation. During RL fine-tuning, the grouped predictor outputs form a multi-property reward that hooks into the LLM for policy updates. \textit{Note :} an in-depth pipeline schematic can be found in Appendix \ref{appendix:detailed_pipeline}.}
\end{figure}

\section{Methodology}

\subsection{Computing physics-grounded refrigerant properties}
\label{section:computing_physics_properties}

Many complex physical properties are hard to measure in practice. Examples include coefficient of performance in vapor compression cycles and global warming potential of refrigerants. To account for data scarcity, our approach consists of two steps: firstly, we train predictors on sub-properties for which annotations are more readily available; secondly, we use established physical models (e.g. equations of state for thermodynamic properties) to compute the final properties.

\paragraph{Coefficient of Performance and Volumetric Concentration} A refrigeration cycle uses the phase changes of a working fluid to transfer heat. The thermodynamic behavior of the molecule in the different steps of the vapor compression cycle is essential as it defines the efficiency of the cycle, called the coefficient of performance (COP). The COP is calculated as the ratio of cooling effect to work input. Another important property which affects the system size is the volumetric cooling capacity ($Q_{vol}$, in $MJ.m^{-3}$), defined as the refrigeration effect per unit volume of refrigerant vapour entering the compressor. For any working pressure and temperature, COP and $Q_{vol}$ are computed as:

\[
COP = \frac{\text{Refrigerating Effect}}{\text{Compressor Work}} = \frac{h_1 - h_4}{h_2 - h_1} \qquad   Q_{vol} = \frac{h_1 - h_4}{v_1}
\]  
Where $h_1$ and $v_1$ are respectively the specific enthalpy and specific volume of the refrigerant at the compressor inlet (state 1), $h_2$ the specific enthalpy at compressor outlet (state  2) and $h_4$ the specific enthalpy of the refrigerant at the evaporator inlet (state 4).

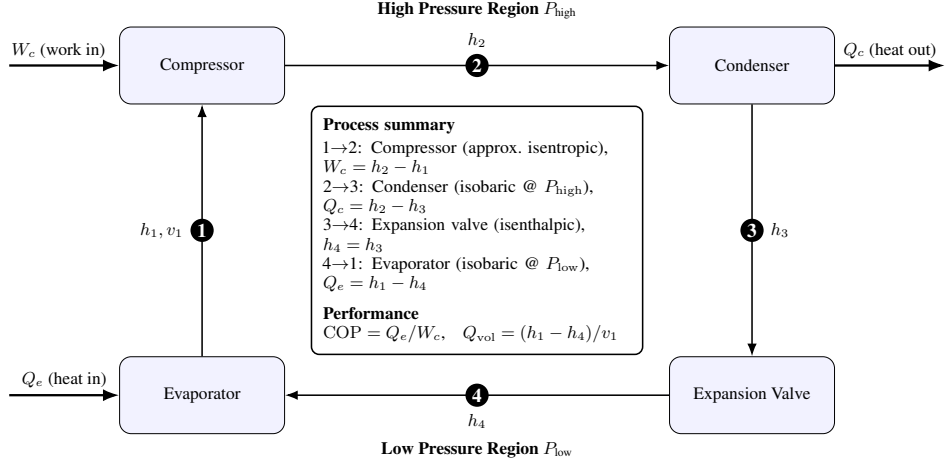
\begin{figure}[t]
    \centering
    \resizebox{0.9\textwidth}{!}{%
    \begin{tikzpicture}[
        font=\footnotesize,
        component/.style={draw, rounded corners=6pt, minimum width=3cm, minimum height=1.4cm, align=center, fill=blue!5},
        flow/.style={thick, -{Latex}},
        heat/.style={very thick, -{Latex}},
        work/.style={very thick, -{Latex}},
        state/.style={circle, fill=black, inner sep=1.2pt}
    ]
    \node[component] (comp) at (-2,6) {Compressor};
    \node[component] (cond) at (8,6) {Condenser};
    \node[component] (evap) at (-2,0) {Evaporator};
    \node[component] (valv) at (8,0) {Expansion Valve};
    
    \node at (3,7) {\textbf{High Pressure Region} $P_{\text{high}}$};
    \node at (3,-1) {\textbf{Low Pressure Region} $P_{\text{low}}$};
    
    \draw[flow] 
      (evap.north) -- 
      node[midway, circle, fill=black, text=white, font=\bfseries, inner sep=1pt, minimum size=12pt] {1}
      node[left=6pt] {$h_1, v_1$}
      (comp.south);
    \draw[flow] 
      (comp.east) -- 
      node[midway, circle, fill=black, text=white, font=\bfseries, inner sep=1pt, minimum size=12pt] {2}
      node[above=6pt] {$h_2$}
      (cond.west);
    
    \draw[flow] 
      (cond.south) -- 
      node[midway, circle, fill=black, text=white, font=\bfseries, inner sep=1pt, minimum size=12pt] {3}
      node[right=6pt] {$h_3$}
      (valv.north);
    
    \draw[flow] 
      (valv.west) -- 
      node[midway, circle, fill=black, text=white, font=\bfseries, inner sep=1pt, minimum size=12pt] {4}
      node[below=6pt] {$h_4$}
      (evap.east);
    
    \draw[heat] 
      ([xshift=-2cm]evap.west) -- (evap.west)
      node[midway, above] {$Q_e$ (heat in)};
    
    \draw[heat] 
      (cond.east) -- ++(2cm,0)
      node[midway, above] {$Q_c$ (heat out)};
    
    \draw[work] 
      ([xshift=-2cm]comp.west) -- (comp.west)
      node[midway, above, xshift=-3pt] {$W_c$ (work in)};
    
    \node[draw, rounded corners=4pt, align=left, text width=5.6cm, fill=white,
          thick, inner sep=6pt] at (3,3) {%
    \textbf{Process summary}\\[2pt]
    1$\to$2: Compressor (approx.\ isentropic), $W_c=h_2-h_1$\\
    2$\to$3: Condenser (isobaric @ $P_{\mathrm{high}}$), $Q_c=h_2-h_3$\\
    3$\to$4: Expansion valve (isenthalpic), \\$h_4=h_3$\\
    4$\to$1: Evaporator (isobaric @ $P_{\mathrm{low}}$), $Q_e=h_1-h_4$\\[6pt]
    \textbf{Performance}\\
    $\mathrm{COP}=Q_e/W_c$,\quad $Q_{\mathrm{vol}}=(h_1-h_4)/v_1$
    };

    \end{tikzpicture}
    }
    \caption{Vapor-compression cycle with four states. 
    Each state is labeled with its specific enthalpy $h_i$, and state~1 also shows the specific volume $v_1$, 
    used in defining the volumetric refrigerating effect $Q_{\text{vol}}$. 
    The coefficient of performance (COP) follows from enthalpy differences between states.}
    \label{fig:vcc_tikz}
\end{figure}

The refrigerant behavior can be modeled with equations of state (EOS) defining the relation between temperature, pressure and enthalpy. We use the modern Peng-Robinson EOS \citep{peng1976new}, which incorporates repulsion and attraction terms, to model the behavior of gases and liquids under different conditions:
\[
P = \frac{RT}{V_m - b} - \frac{a \alpha(T)}{V_m (V_m + b) + b (V_m - b)}.
\]

where \( P \) is pressure, \( V_m \) is molar volume, \( T \) is temperature, and \( R \) is the ideal gas constant. The parameters \( a \), \( b \), and the temperature-dependent factor \( \alpha(T) \) are expressed in terms of critical properties and the acentric factor, which needs to be given as input to the model:
\[
b = 0.07780 \, \frac{RT_c}{P_c}, \quad a = 0.45724 \, \frac{R^2 T_c^2}{P_c}, \quad \alpha(T) = \left[1 + m \left(1 - \sqrt{T/T_c} \right) \right]^2,
\]
with
\[
m = 0.37464 + 1.54226 \, \omega - 0.26992 \, \omega^2.
\]

To simulate the thermodynamic cycle of a molecule and COP and $Q_{vol}$ calculations for any working pressures and temperatures, the relationships between real refrigerant enthalpy, entropy and other molecular properties such as pressure and temperature have to be computed. Ideal gas properties are first computed through NASA polynomials, which ensure proper thermodynamic relationships between heat capacity ($C_p$), enthalpy ($H$) and entropy ($S$) before using the EOS to obtain departure functions from ideal to real behaviour. Detailed explanations are in appendix \ref{appendix:cop_simulation}.

\paragraph{Global Warming Potential} The second characteristic to consider is the environmental impact, predominantly measured by the global warming potential (GWP). The GWP$_{100}$ quantifies the cumulative radiative forcing impact of a greenhouse gas (GHG) relative to CO$_2$, over a 100-year time horizon :
\begin{equation*}
    \text{GWP}_{100} = \frac{\text{AGWP}_{\text{X}}(100)}{\text{AGWP}_{\text{CO}_2}(100)}
\end{equation*}
where $\text{AGWP}_{\text{X}}(100)$ is the Absolute Global Warming Potential of species X, and $\text{AGWP}_{\text{CO}_2}(100) = 9.0 \times 10^{-14}\ \text{W\,m}^{-2}\,\text{yr\,kg}^{-1}$ is the same metric for CO$_2$, used as a reference.

AGWP$_\text{X}$, as defined in the IPCC reports and detailed in \citet{Zieger2025ClimateCM} combines the atmospheric lifetime of a molecule and its radiative efficiency :
\begin{equation*}
    \text{AGWP}_{\text{X}}(100) = \text{RE}_{\text{X}} \times \tau_{\text{X}} \times \left(1 - e^{-100/\tau_{\text{X}}} \right)
\end{equation*}
Here, $\text{RE}_{\text{X}}$ is the \textbf{radiative efficiency} of X (W\,m$^{-2}$\,kg$^{-1}$), i.e., the instantaneous radiative forcing (net change in radiative flux at the top of the atmosphere caused by the presence or increase of a greenhouse gas) per unit mass increase. $\tau_{\text{X}}$ is the \textbf{atmospheric lifetime} of X in years. The atmospheric lifetime $\tau$ is the mean residence time of a molecule before it is removed via chemical reactions or photolysis.  As shown in \citet{Kazakov2012ComputationalDO}, the accuracy in estimating  $\tau_{\text{X}}$ is important mainly for the cases when  $\tau_{\text{X}} \ll 100 $. For this range of lifetimes, reaction with OH appears to be the dominant loss mechanism. The corresponding lifetime is:
\begin{equation*}
    \tau_{\text{X}} = \frac{1}{k_{{OH}}[{OH}]}
\end{equation*}
where $k_{{OH}}$ is the rate constant (cm$^3$\,molecule$^{-1}$\,s$^{-1}$), and $[{OH}] \approx 1 \times 10^6$ molecules\,cm$^{-3}$ \citep{Kazakov2012ComputationalDO}.

\paragraph{Flammability} Although flammability requirements are being reconsidered to help find new refrigerants, many safety codes like \citet{ASHRAE34_2022} still require nonflammable refrigerants. The ASHRAE documentation uses the Lower Flammability Limit (LFL) defined as the minimum concentration capable of propagating a flame through a homogeneous mixture of refrigerant and air at standard conditions (23°C and 101.3 kPa), typically expressed in vol\% or $kg.m^{-3}$.

\subsection{Predicting refrigerant properties from SMILES}
\label{section:property_predictors}

Building robust property predictors is crucial to guide the search of molecules during post-training RL towards desired tradeoffs. As annotated datasets of thermodynamic property-measurements are scarce, we leverage laws introduced in section \ref{section:computing_physics_properties} for more accurate scoring. Our property predictors, similarly to our generative backbone, are sequence models that encode \textit{SMILES} (Simplified Molecular Input Line Entry System, \citet{Weininger1988}): textual representations of molecules in ASCII characters, representing stereochemistry, cycles and branches. A SMILES string is computed by traversing the molecular graph, and is therefore not unique, enabling augmentation. We base our predictors on the \textit{SMIles Transformer Encoder Decoder (SMI-TED)} \citep{Soares2025}, in order to leverage its pre-trained dense embeddings and to avoid handcrafted features. 


\subsection{\texttt{RefGen} generative model training}

Our model is based on Llama 3.2 1B \citep{llama3herd2024}, an open and highly capable LLM for instruction-conditioned tasks. We adopt the same tokenization scheme, Byte-Pair Encoding, which we found to work well with SMILES. We fine-tune (SFT) the pre-trained Llama model on unconditioned SMILES generation; re-using this LLM allows to inherit instruction-tuning and language understanding capabilities for downstream chemical reasoning tasks. In a subsequent post-training phase (RLFT), the model learns to generate molecular structures that more likely satisfy the acceptance ranges for thermodynamic properties.

\paragraph{Supervised Fine-Tuning (SFT)} We fine-tune the model on standard causal language modeling on sequences $\mathbf{x} \in \mathcal{D}_{SFT} \,\,,\,\, \mathbf{x} = (x_1, x_2, \ldots, x_T)$ where $x_t \in \mathcal{V}$ denotes tokens from the vocabulary $\mathcal{V}$ and $T$ is the sequence length. Here, $\mathcal{D}_{SFT}$ is a combination of Pubchem \citep{kim2021pubchem}, ChEMBL \citep{chembl} and SureChEMBL \citep{papadatos2016surechembl} processed with our filtering pipeline, detailed in Appendix \ref{sft:llama}, counting a total of $\sim 37$M sequences ($\sim1.5$B tokens). A format for structured SMILES generation is adopted: sequences are delimited with XML-style tokens \texttt{<s>} or \texttt{<smiles>} and \texttt{</s>} or \texttt{</smiles>}, e.g. \texttt{<s>CC1(F)C(F)C1(F)F</s>}. \label{section:sft_step}

\paragraph{RL Finetuning (RLFT)} We consider Group-Relative Policy optimization (GRPO) \citep{shao2024deepseekmath}, which introduces a simplification of the PPO objective based on advantage estimation over multiple rollouts, given (multiple) callable reward functions. We can thus consider the property predictor networks from \ref{section:property_predictors} as scoring functions, each depending on the defined acceptance region per thermodynamic property, e.g. number of atoms between 7 and 18, further details in Appendix \ref{appendix:grpo_details}. \label{section:rlft_step}

\section{Experimental setup}
\subsection{Refrigerant property prediction}

We compile datasets with annotated properties for SMILES structures: \texttt{Tc}, \texttt{Pc} and \texttt{w} are extracted from \citet{bell_chemicals_2016_2024}; NASA polynomials for the enthalpy and entropy from \cite{Farina_Jr_2021}; radiative efficiencies from \citet{Muthiah2023DevelopingML};  \texttt{k(OH)} reaction constants from \citet{McGillen2020DatabaseFT}; \texttt{LFL} values are compiled from \citet{Maury2023} and \citet{bell_chemicals_2016_2024}. We fine-tune SMI-TED end-to-end on our property-measurement datasets. We apply SMILES augmentation by canonicalizing the structures in the dataset and then by generating multiple graph traversals with \texttt{RDKIT}'s \texttt{MolToSmiles} functionality. Copies of the base SMI-TED model are fine-tuned for the various properties with train/valid/test splits. Additionally, we test our \texttt{COP} predictor on a \texttt{CoolProp} \citep{CoolProp2025} test split, which also implements detailed EOS for many refrigerant compounds. Accuracy of the \texttt{GWP} predictor is estimated on data from \citet{EPA_IPCC_GWPs_2023}. For the reaction rate with the OH radical \texttt{k(OH)}, we additionally compare performance with a group contribution method based on \citet{Kwok1995Estimation} and open source notes from the EPA AOPWIN software \citep{US_EPA_EPI_2012}. Further details in Appendix \ref{appendix:prop_pred_choice} and \ref{appendix:prop_pred_training_details}.


\subsection{Reinforcing optimal refrigerant molecules}

To sample optimal refrigerant molecules wrt. constraints on thermodynamic properties, we fine-tune our model with GRPO on completions of \texttt{<s>} (unconditioned molecule generation). The reward signal consists of a linear combination of five scoring functions with relative weights obtained via hyperparameter search: 
$$
    R_{\text{total}}(\mathbf{x}) = R_{diversity}(\mathbf{x}) \cdot \sum_{i=1}^k w_i R_i(\mathbf{x}) 
$$
where $w_i$ represents the relative reward weight, $R_i(\mathbf{x})$ is the reward from the $i$-th property predictor, and $k$ is the number of objectives. We empirically observe the properties \texttt{COP}, $Q_{vol}$, $T_c$ to be the most impactful for convergence, hence we allocate more weight. A diversity reward is considered as global scaling factor to ideally re-weight good molecules by novelty. Reward functions and training details are provided in Appendix \ref{appendix:grpo_details}.

\begin{figure}[h!]
\centering
\begin{minipage}{0.4\textwidth}
    \centering
    \begin{tabular}{lc}
        \toprule
        \textbf{Reward} & \textbf{$w_i$} \\
        \midrule
        COP \& $Q_{vol}$ & 0.40 \\
        $T_c$ & 0.40 \\
        Molecular length & 0.10 \\
        GWP & 0.05 \\
        LFL & 0.05 \\
        \bottomrule
    \end{tabular}
    \caption{Reward weights}
    \label{tab:reward_weights}
\end{minipage}%
\hfill
\begin{minipage}{0.6\textwidth}
    \centering
    \includegraphics[width=\linewidth]{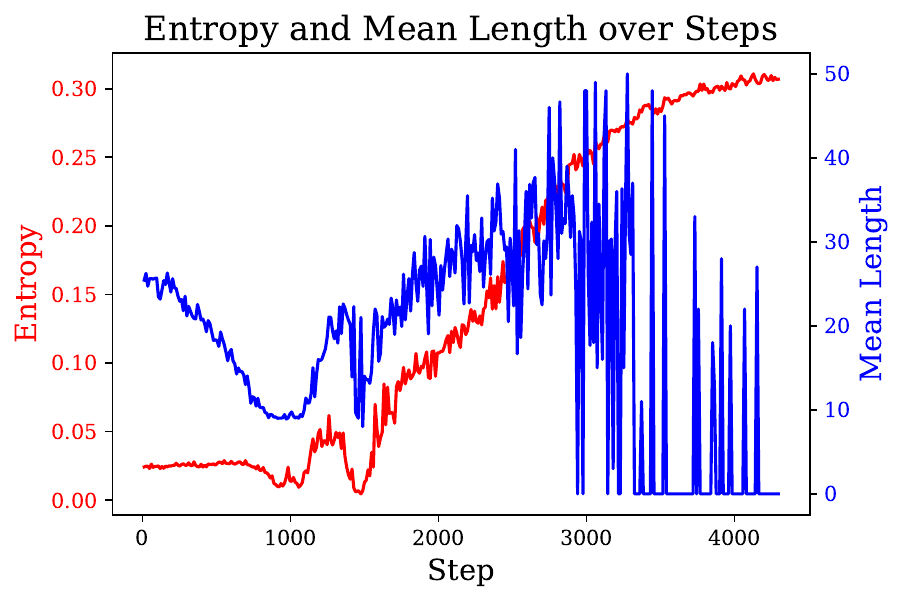}
    \caption{Mode collapse with entropy maximization.}
    \label{fig:entropy_mean_length}
\end{minipage}
\end{figure}

 To encourage diversity, we test two approaches: (1) introducing an entropy regularizer, in spirit of \cite{eysenbach2022maximumentropyrlprovably}; (2) using a diversity and repetition reward signal as in \cite{Loeffler2024Reinvent4}. We find the latter to better prevent mode collapse (an instance is displayed in Figure \ref{fig:entropy_mean_length}) and to yield more stable training dynamics, as entropy maximization clashes with the idea of narrowing down search to a subset of optimal molecules, thus requiring additional techniques such as annealing.

\section{Results}

We evaluate two main components of the \texttt{Refgen} pipeline: we first assess the property predictor's coverage across the molecular space; and subsequently evaluate our generative model to quantify, within the distribution covered by our predictors, the quality of generated structures.

\begin{figure}[h!]
    \centering
    \begin{minipage}{0.49\textwidth}
        \centering
        \includegraphics[width=\linewidth]{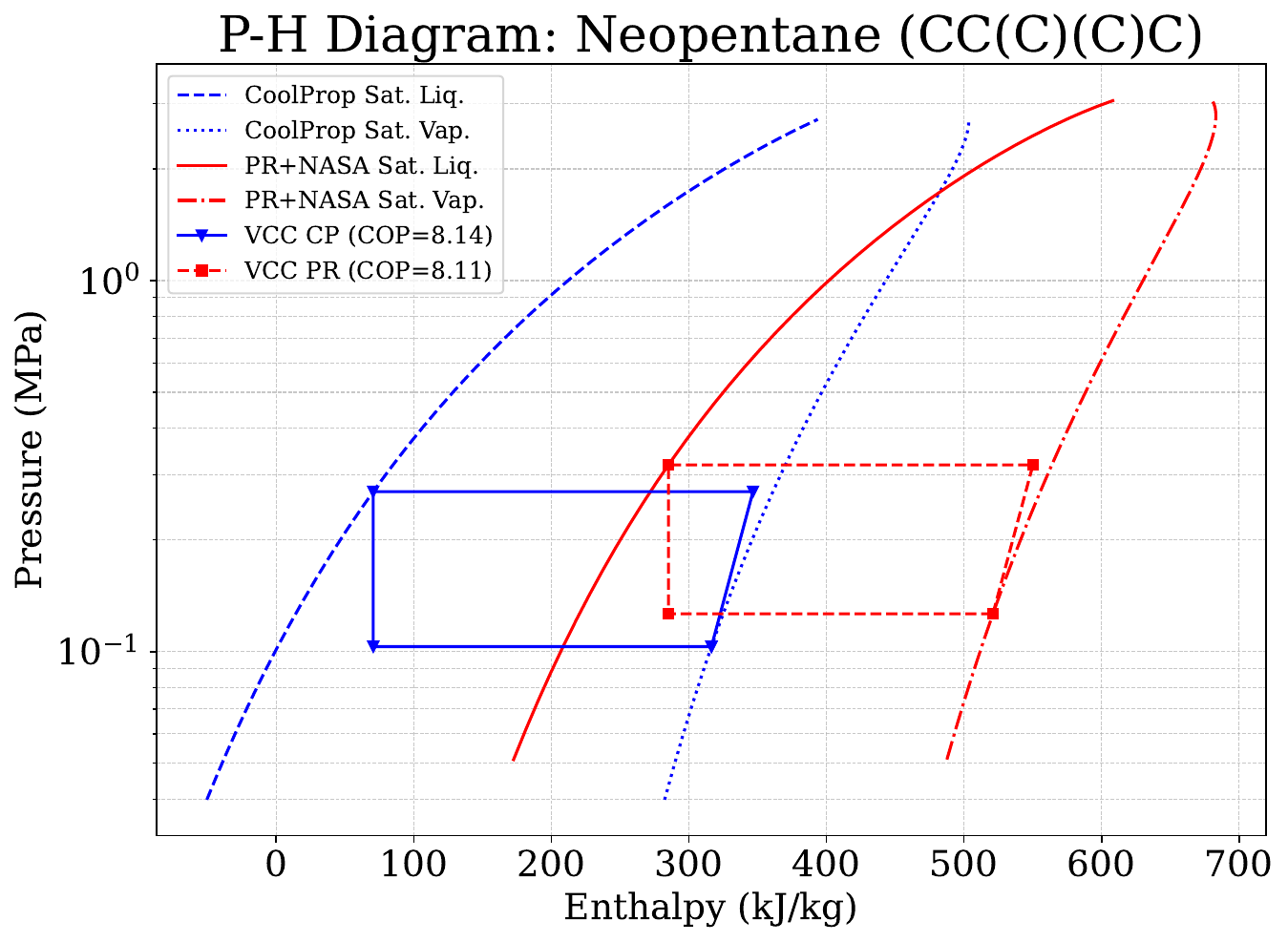}
        \label{fig:coolprop_neopentane}
    \end{minipage}%
    \hfill
    \begin{minipage}{0.49\textwidth}
        \vspace{-11pt}
        \centering
        \includegraphics[width=\linewidth]{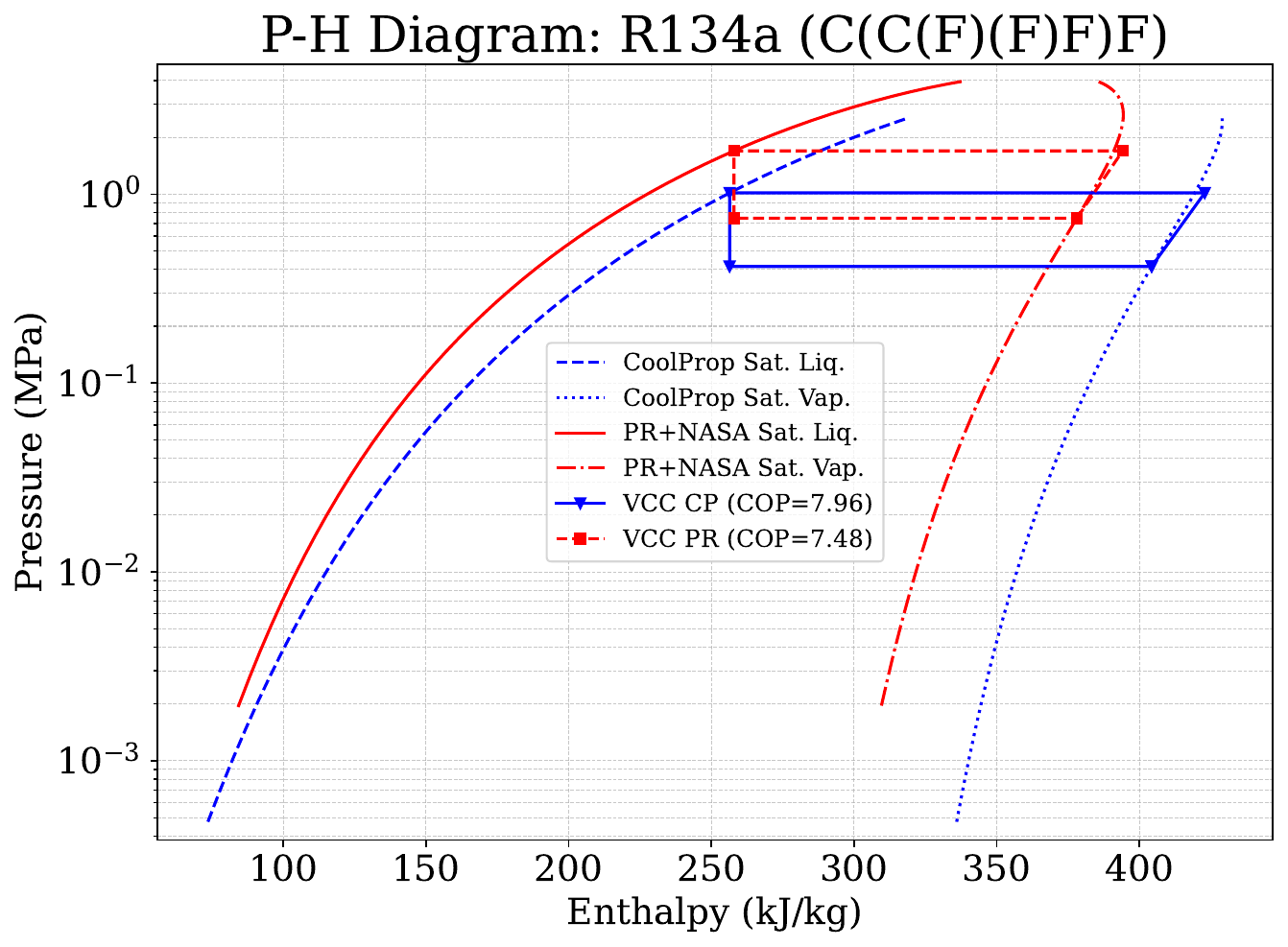}
    \end{minipage}
    \caption{Comparison between ground-truth saturation dome from CoolProp (blue) and predicted (red) for molecules Neopentane (left) and R134a (right). The COP is computed for condenser and evaporator temperatures at 10°C and 40°C respectively. Dome translation shifts are due to varying reference points in the enthalpy and they are irrelevant for the computation of COP or $Q_{vol}$.}
    \label{fig:coolprop_comparison}
\end{figure}

\paragraph{Property predictor performance} We first ground the choice of the SMI-TED backbone for property prediction with a comparison to a gradient-boosting baseline with handcrafted features. Experimental results are reported in Table \ref{tab:mean_performance_2}, Appendix \ref{appendix:prop_pred_choice}. We find the features learned with SMI-TED to be robust and SMILES augmentation proves beneficial, as redundant representations are mapped, leading to increased test accuracy overall. For the fine-tuned SMI-TED models, we observe low prediction errors in and out of distribution: for \texttt{COP}, we compute the saturation dome reconstruction accuracy by comparing the saturation enthalpy difference RMSE at sampled critical temperatures along the dome between the predicted cycle and CoolProp's very accurate EOS considered as ground-truth. We also compute COP mean absolute error for operating temperatures of 10°C and 40°C (Figure \ref{fig:coolprop_comparison} and Table \ref{tab:coolprop_system_metrics}). Similarly, we report, for the \texttt{GWP} property, accurate distribution coverage, with the \texttt{k(OH)} group contribution method showcasing the best generalization ability; results for all the predictors are reported in Appendix \ref{appendix:prop_pred_results}.

\begin{table}[h!]
    \centering
    \caption{Median property scores\tablefootnote{The COP comparison cannot be accurately made as only 22 molecules had properties in range for which a saturation dome and therefore a COP and $Q_{vol}$ could be computed. Concerning the GWP, comparison is difficult as GWP predictions for these complex molecules is certainly out of distribution} for 1000 unique molecules sampled from the base (\texttt{Refgen}-SFT) and fine-tuned (\texttt{Refgen}-RLFT) models .}
    \vspace{2pt}
    \label{tab:distrib_analysis}
    \begin{tabular}{l c c c c c c}
        \toprule
        \textbf{Setup} & $T_c$ [K] & $N_{\text{atoms}}$ & GWP & LFL [kg/m\textsuperscript{3}] & COP & $Q_{vol}$[MJ/m\textsuperscript{3}]\\
        \midrule
        Base Model & $841$ & $22$ & $0.04$ & $0.07$ & $8.66$ & $0.002$ \\
        \textbf{Finetuned Model} & $\mathbf{403}$ & $\mathbf{9}$ & $\mathbf{13}$ & $\mathbf{0.61}$ & $\mathbf{7.99}$ & $\mathbf{1.179}$ \\
        \bottomrule
    \end{tabular}
\end{table}

\begin{figure}[h!]
    \centering
    \includegraphics[width=1.0\linewidth]{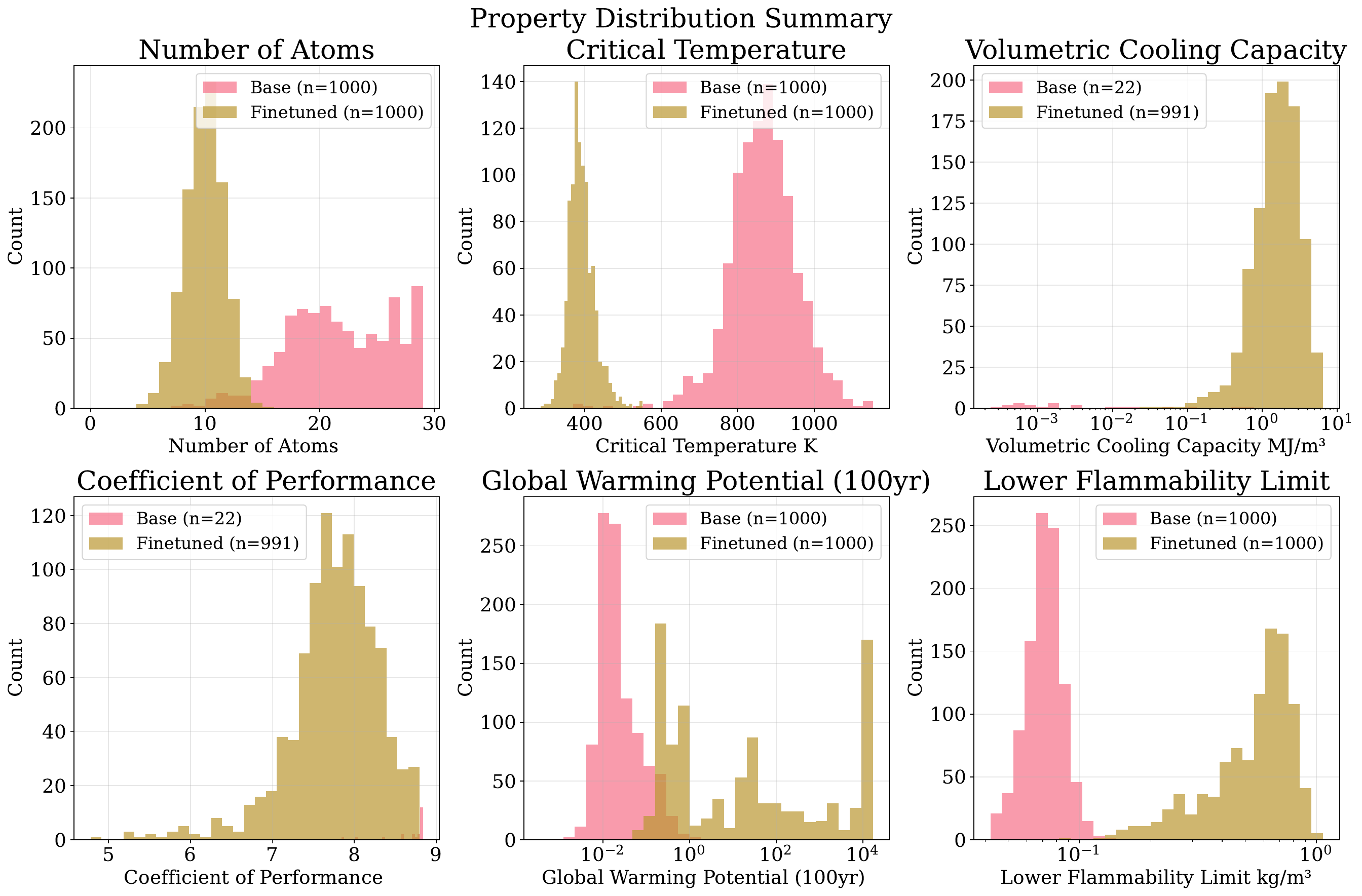}
    \caption{Comparison between distribution plots for our base (\texttt{Refgen}-SFT) and finetuned (\texttt{Refgen}-RLFT) models in terms of length, \texttt{Tc}, \texttt{Q}$_{vol}$, \texttt{COP}, \texttt{GWP}, \texttt{LFL}.}
    \label{fig:distrib_analysis}
\end{figure}

\paragraph{De Novo molecular generation} We first illustrate the difference between the distribution of generated structures after SFT (base model with no specialization) and after RLFT (fine-tuned model with property optimization) (Table \ref{tab:distrib_analysis}, Figure \ref{fig:distrib_analysis}). Optimizing for thermodynamic properties significantly changes the target region towards small fluorinated compounds such as HFC's when optimizing for $T_c$ and molecular length, or towards HFO's when adding GWP and flammability constraints. Additionally optimizing for $Q_{vol}$ explores new regions of the chemical space presenting multiple novel SMILES not listed in \citet{McLinden2017}.

\begin{figure}[h!]
    \centering
    \includegraphics[width=1\linewidth]{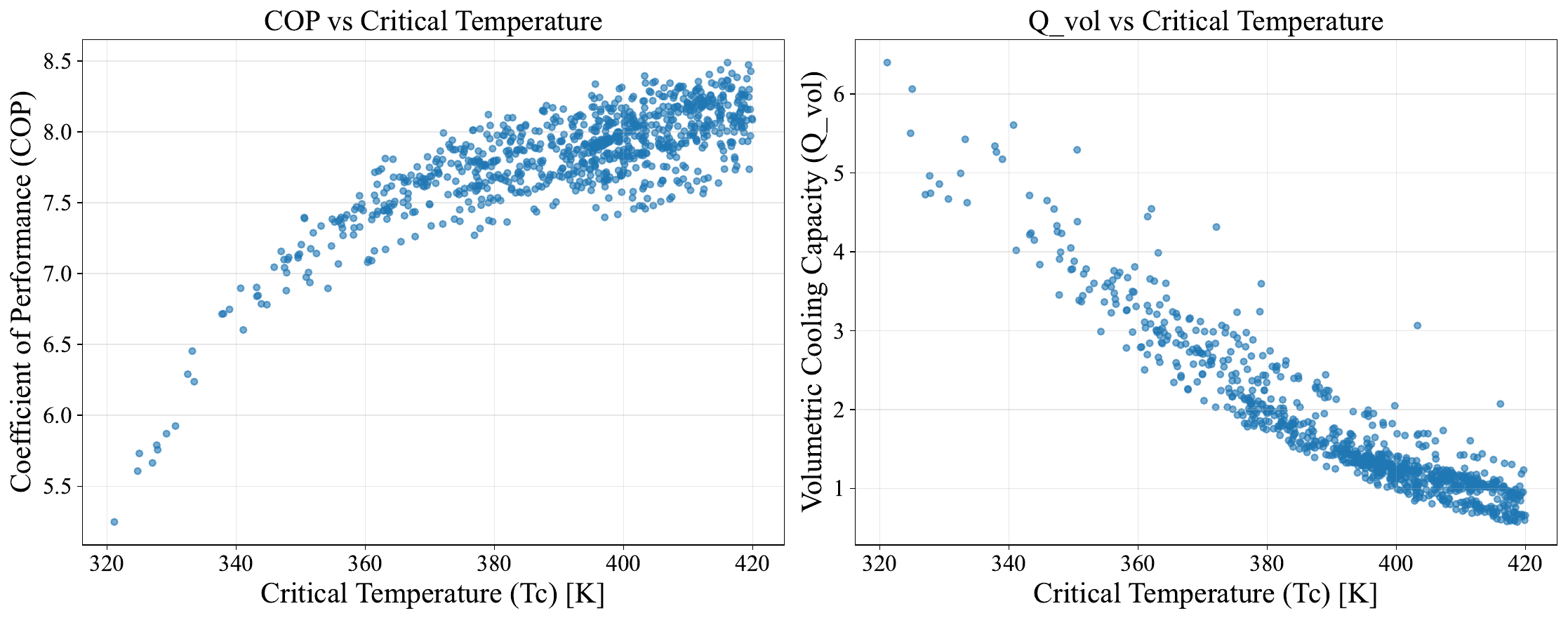}
    \caption{Comparative plots for \texttt{COP} to \texttt{Tc} and $Q_{vol}$ to \texttt{Tc} for compounds generated with \texttt{Refgen}.}
    \label{fig:tc_qvol_tradeoff}
\end{figure}

\begin{figure}[h!]
    \centering
    \includegraphics[width=0.9\linewidth]{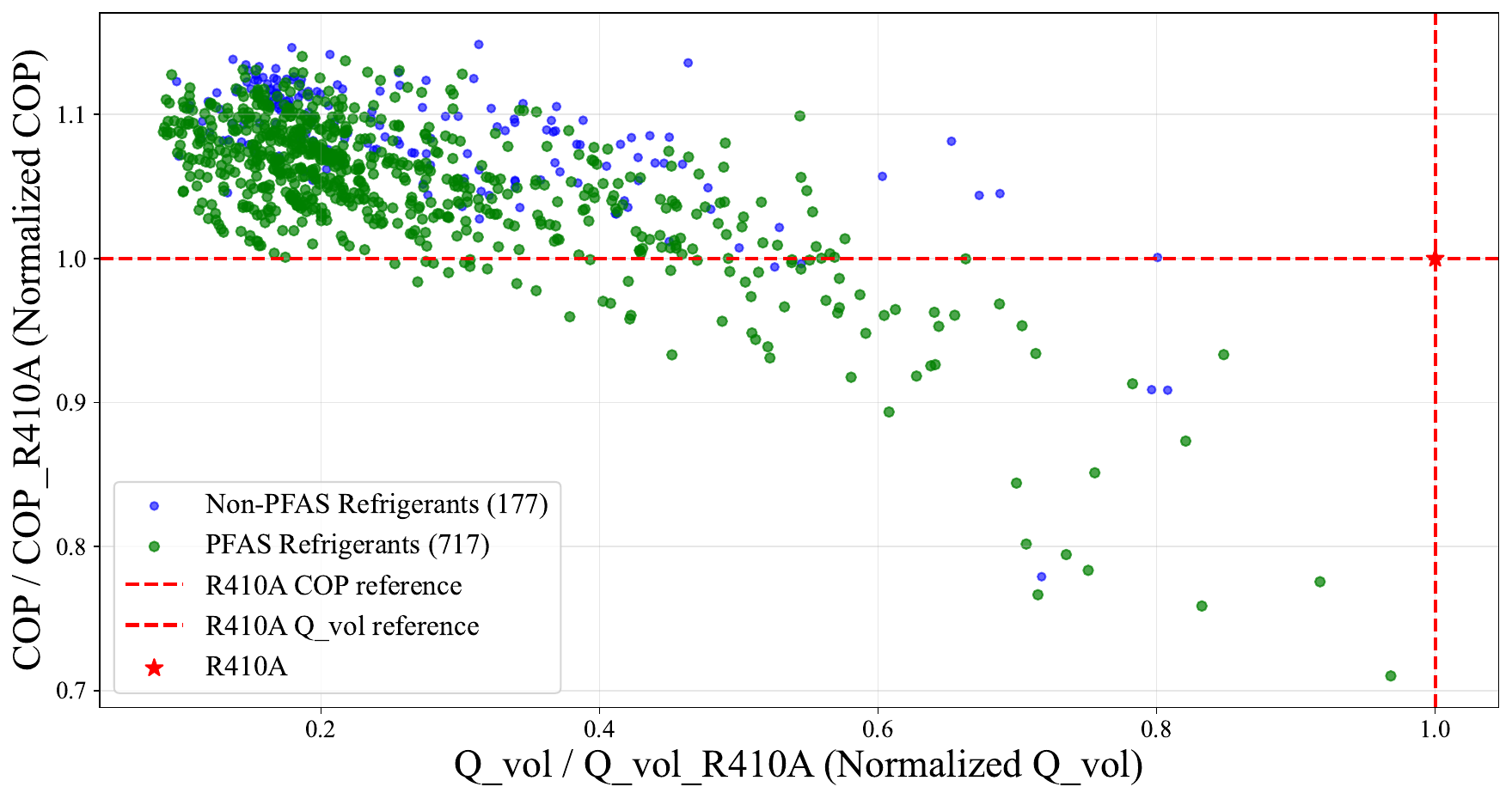}
    \caption{Generated molecules visualized by \texttt{COP} to $Q_{vol}$ ratio, compared to R-410A reference.}
    \label{fig:comparison_q_vol}
\end{figure}

\label{sec:filtering_constraints}
Finally, we consider a set of constraints to select optimal candidates based on principles from \citet{McLinden2017}: \texttt{COP > 5} ; \texttt{320K < Tc < 420K} ; \texttt{LFL > 0.1} $kg.m^{-3}$ ; we also impose \texttt{GWP < 10}, which is lower than all existing regulations but still keeping some margin compared to \texttt{GWP = 1} in $CO_2$. We filter out \texttt{=CF2} and \texttt{-OF} groups according to stability and toxicity warnings detailed in \citet{McLinden2017}. Generated molecules reveal a tradeoff between \texttt{COP} and $Q_{vol}$ depending on critical temperature, Figure \ref{fig:tc_qvol_tradeoff}: the higher the \texttt{Tc}, the higher the \texttt{COP}, but the lower the $Q_{vol}$. This behaviour, observed in the reference work of \citet{McLinden2017} confirms the importance of our physics-grounded approach, which ensures thermodynamically sound predictions. In Figure \ref{fig:comparison_q_vol}, we compare the \texttt{COP} to $Q_{vol}$ tradeoff of the best molecule candidates with respect to R-410A (\texttt{COP = 7.39}, $Q_{vol}$ \texttt{= 6.61}) as in the reference. R-410A is a heavily used refrigerant blend that is considered to have some of the best thermodynamic properties in the vapour compression cycle; we observe similar or competitive tradeoffs for our compounds, which additionally optimize for diversity and other properties introduced in our new pipeline. 

Finally, we screen out per- and polyfluorinated substances (\texttt{PFAS}) according to the OECD definition \citep{oecd2021reconciling} for potential environmental risk of toxicity. We observe  $\sim$20\% of generated molecules to be \texttt{non-PFAS}, despite no constraint having been added.

\begin{figure}[h!]
\centering

\begin{minipage}{0.6\textwidth}
    \resizebox{\linewidth}{!}{
        \begin{tabular}{lrrrrrr}
            \toprule
            \textbf{SMILES} & \textbf{COP} & \textbf{$T_c$ (K)} & \textbf{GWP100} & \textbf{LFL (kg/m$^3$)} & \textbf{$Q_{vol}$} \\
            \midrule
            \texttt{C(F)N(F)N(F)F} & 6.71 & 337.82 & 3.95 & 0.52 & 5.34 \\
            \addlinespace
            \texttt{N(F)C(F)N(F)F} & 6.72 & 338.05 & 0.25 & 0.54 & 5.26 \\
            \bottomrule
        \end{tabular}
    }
\end{minipage}
\hfill 
\begin{minipage}{0.35\textwidth}
    \includegraphics[width=2.2cm]{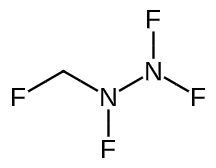}
    \includegraphics[width=2.2cm]{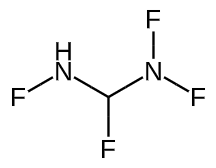}
\end{minipage}

\caption{Example refrigerant candidates with properties and 2D structures, full list in Appendix \ref{appendix:final_molgen_results}.}
\label{fig:table_and_structures}
\end{figure}

\section{Discussion and Conclusion}

We introduced \texttt{Refgen}, a novel generative pipeline for refrigerant molecules which successfully handles thermodynamic and environmental tradeoffs. Our model re-discovers HFCs and HFOs when appropriate constraints are imposed, while also proposing novel classes of candidates with the provided tradeoffs. Further work is currently focused on scaling up constraints, increasing generation diversity, and most importantly, verifying the validity of sampled candidates in a real lab.

To improve the overall quality of molecules, additional properties could be taken into account such as toxicity or molecular stability. Toxicity prediction highly differs in refrigerant discovery with respect to drug design. In the first case, long term exposures defined by workplace security panels usually define toxicity levels on an individual molecular basis with no clearly defined formula, making estimation very difficult. Molecular stability is equally relevant, as some of the generated candidates contain highly strained 3-atom cycles, which are known to be unstable. 

\begin{minipage}[t]{0.55\textwidth} 
    As we sample more sequences from our generative model, we start to observe a saturation in the number of candidates satisfying our constraints, Table \ref{tab:filtered_mols_scaling}. This suggests potential limits in exploration and diversity. Further work includes investigating novel generative approaches and techniques to emphasize variety.
\end{minipage}%
\hfill
\begin{minipage}[t]{0.4\textwidth}
    \vspace{-10pt}
    \centering
    \captionof{table}{Selected unique molecules by constraints satisfaction, Section \ref{sec:filtering_constraints}.}
    \begin{tabular}{cc}
        \toprule
        $N_{\text{generated}}$ & $N_{\text{filtered}}$ \\
        \midrule
        65 536 & 786 \\
        1 024 000 & 894 \\
        \bottomrule
    \end{tabular}
    \label{tab:filtered_mols_scaling}
\end{minipage}

We aim at integrating our \textit{in-silico} pipeline in the real development process, where candidates sampled from our generative model can be further validated in lab. On this line of research, we seek for experimentation in collaboration with chemists and HVAC experts to assess refrigerant synthesizability, stability, environmental impact and practical efficacy in vapor compression cycles.

\newpage
\bibliographystyle{plainnat}
\bibliography{references}


\appendix
\newpage

\section{Detailed \texttt{Refgen} pipeline}
\label{appendix:detailed_pipeline}

\begin{figure}[h!]
    \centering
    \includegraphics[width=1.48\linewidth, angle=-90]{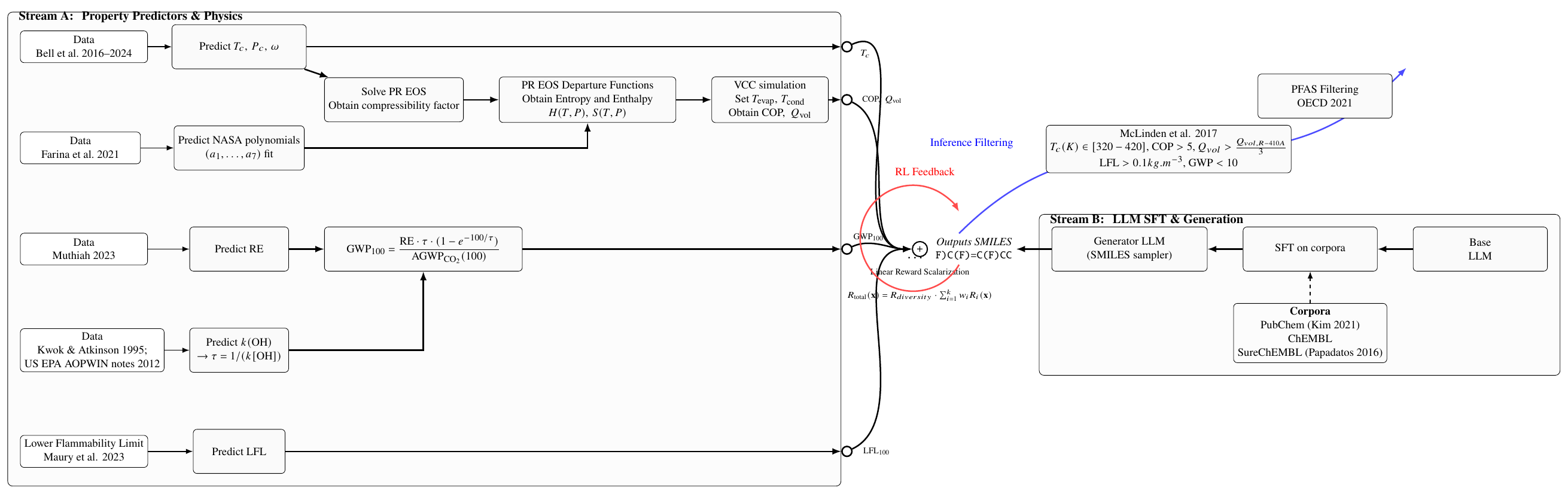}
    \caption{Detailed \texttt{Refgen} pipeline with property prediction details and datasets (Stream A, left) as well as SMILES generation (Stream B, left).}
    \label{fig:placeholder}
\end{figure}
\newpage

\section{Related Work}
\label{appendix:related_work}
While traditional screening is limited to known compounds from databases, recent \textit{generative modeling} techniques have shown an ability to explore the chemical space, allowing for the generation of new ('\textit{de novo}') molecular structures while balancing different property tradeoffs, constraining the molecular space towards a specific region of interest. Methods vary by design choices: using graph representations \citep{han2021reliablegraphneuralnetworks} or textual e.g. SMILES \citep{Weininger1988}; using sequence models \citep{ross2022large, Loeffler2024Reinvent4}, GFlowNets \citep{jain2023multi} or discrete diffusion \citep{tang2025peptune, vignac2023digress} and flow matching \cite{dunn2024exploringdiscreteflowmatching, cremer2025flowrflowmatchingstructureaware}. Recently sequence models, usually in the context of \textit{Large Language Models} (LLMs), gained traction in the molecule discovery community due to their scalability and versatility, allowing to learn from collections of billions of compounds e.g. ZINC \cite{tingle2023zinc22}. Post-training fine-tuning techniques for LLMs, such as as Direct Preference Optimization (DPO) \citep{rafailov2023direct} or Group Relative Policy Optimization (GRPO) \citep{shao2024deepseekmath}, well transferred from the natural language processing community to chemistry and computational biology, allowing for the development of chemical reasoning models \cite{bran2023chemcrowaugmentinglargelanguagemodels, narayanan2025trainingscientificreasoningmodel} or models for molecular structures \cite{zholus2024bindgptscalableframework3d, ross2022large} and proteins \cite{hesslow2022ritastudyscalinggenerative}.

\section{COP calculation and cycle simulation}
\label{appendix:cop_simulation}

\subsection{The Peng-Robinson Equation of State}

To model the behavior of gases and liquids under different conditions, we use an equation of state. While the simplest EOS is usually the \emph{ideal gas law}, more refined models, such as the \emph{Peng--Robinson (PR) equation of state}, incorporate repulsion and attraction between molecules. Its form is:
\[
P = \frac{RT}{V_m - b} - \frac{a \alpha(T)}{V_m (V_m + b) + b (V_m - b)}.
\]

where \( P \) is pressure, \( V_m \) is molar volume, \( T \) is temperature, and \( R \) is the ideal gas constant. The parameters \( a \), \( b \), and the temperature-dependent factor \( \alpha(T) \) are expressed in terms of critical properties and the acentric factor:
\[
b = 0.07780 \frac{RT_c}{P_c}, \quad a = 0.45724 \frac{R^2 T_c^2}{P_c}, \quad \alpha(T) = \left[1 + m \left(1 - \sqrt{T/T_c} \right) \right]^2,
\]
with
\[
m = 0.37464 + 1.54226\omega - 0.26992\omega^2.
\]
Importantly, these coefficients depend on the critical temperature $T_c$, critical pressure $P_c$ as well as the accentric factor $w$, which need to be inputs to the model. 

To quantify how real fluids deviate from ideal behavior, we define the \emph{compressibility factor}:
\[
Z = \frac{P V_m}{RT}.
\]
For an ideal gas, \( Z = 1 \). Deviations from unity reflect the presence of intermolecular forces: \( Z < 1 \) indicates attraction, while \( Z > 1 \) suggests repulsion or volume exclusion.

The Peng Robinson EOS can be rewritten as a third degree polynomial in the compressibility factor Z. 
\[
Z^3 - (1 - B)Z^2 + (A - 3B^2 - 2B)Z - (AB - B^2 - B^3) = 0
\]

Where 
\[
A = \frac{a \alpha P}{(RT)^2} \quad \text{and} \quad B = \frac{bP}{RT}
\]

The roots of this equation are key to obtaining thermodynamic relationships at the two phase region between liquid and vapor, and obtaining the COP.  Solving the equation can lead to two scenarios : 
\begin{itemize}
    \item In the two phase region : 3 real roots 
    \begin{itemize}
        \item The smallest root : $Z_{liq}$ of the liquid phase
        \item The largest root : $Z_{vap}$ of the vapor phase
        \item The middle root is an artifact with no thermodynamic significance
    \end{itemize}
    \item Outside the two phase region : 1 real root
\end{itemize}

\subsection{Ideal to Real fluid and gas behaviour with Nasa Polynomials}

\subsubsection{Enthalpy, Entropy and Heat Capacity}

To interpret thermodynamic models, we must define three key quantities: enthalpy, entropy, and heat capacity.

The \emph{enthalpy} \( H \) includes internal energy and the energy required to occupy volume:
\[
H = U + PV.
\]
It reflects the heat required to raise temperature at constant pressure.

The \emph{entropy} \( S \) quantifies the degree of disorder or number of accessible microstates. Its temperature derivative is related to the heat capacity:
\[
\left( \frac{\partial S}{\partial T} \right)_P = \frac{C_p}{T}.
\]

The \emph{heat capacity at constant pressure} is given by
\[
C_p = \left( \frac{\partial H}{\partial T} \right)_P.
\]

\subsubsection{Nasa Polynomial thermodynamic representations}

NASA polynomials address the challenge of consistently calculating heat capacity, enthalpy, and entropy while maintaining computational efficiency. 

Given the heat capacity $C_p(T)$ as a function of temperature $T$, the thermodynamic relationships enable us to derive the enthalpy $H(T)$ and entropy $S(T)$ through integration with respect to temperature:

\[
H(T) - H(T_{\text{ref}}) = \int_{T_{\text{ref}}}^{T} C_p(\tau) d\tau
\]
\[
S(T) - S(T_{\text{ref}}) = \int_{T_{\text{ref}}}^{T} \frac{C_p(\tau)}{\tau} d\tau
\]

where $T_{\text{ref}}$ is a reference temperature at which enthalpy and entropy are defined relative to some standard state.

NASA uses a 4th-degree polynomial representation for the dimensionless heat capacity: 
\[
C_p(T)/R = a_0 + a_1T + a_2T^2 + a_3T^3 + a_4T^4
\]
where $R$ is the ideal gas constant and $a_i$ are the polynomial coefficients. For this form, the indefinite integrals required for enthalpy and entropy can be calculated analytically:
\[
\int \frac{C_p(T)}{R} dT = a_0T + \frac{a_1}{2}T^2 + \frac{a_2}{3}T^3 + \frac{a_3}{4}T^4 + \frac{a_4}{5}T^5 + \text{constant}_H
\]
and similarly for the entropy:
\[
\int \frac{C_p(T)}{RT} dT = \int \left(\frac{a_0}{T} + a_1 + a_2T + a_3T^2 + a_4T^3\right) dT = a_0\ln(T) + a_1T + \frac{a_2}{2}T^2 + \frac{a_3}{3}T^3 + \frac{a_4}{4}T^4 + \text{constant}_S
\]

A key property that arises from these analytical expressions is the shared coefficients across all three thermodynamic quantities. This is a form of "parameter sharing" ensures the desired consistency. The resulting representations for heat capacity, enthalpy, and entropy in the NASA 7-coefficient format are:
\[
C_p(T) / R = a_0 + a_1T + a_2T^2 + a_3T^3 + a_4T^4
\]
\[
H(T) / RT = a_0 + \frac{a_1}{2}T + \frac{a_2}{3}T^2 + \frac{a_3}{4}T^3 + \frac{a_4}{5}T^4 + \frac{a_5}{T} 
\]
\[
S(T) / R = a_0 \ln(T) + a_1T + \frac{a_2}{2}T^2 + \frac{a_3}{3}T^3 + \frac{a_4}{4}T^4 + a_6
\]

The coefficients 
\[
a_5 = -H(T_{\text{ref}})/(RT_{\text{ref}})
\]
and
\[
a_6 = S(T_{\text{ref}})/R - a_0\ln(T_{\text{ref}}) - a_1T_{\text{ref}} - \frac{a_2}{2}T_{\text{ref}}^2 - \frac{a_3}{3}T_{\text{ref}}^3 - \frac{a_4}{4}T_{\text{ref}}^4
\]
incorporate the integration constants, setting the reference enthalpy and entropy values at a chosen reference temperature $T_{\text{ref}}$ (often 298.15 K). By defining $H(T)$ and $S(T)$ through the analytical integrals of the same $C_p(T)$ polynomial, the fundamental thermodynamic relationships between $C_p$, $H$ and $S$ are inherently satisfied.

\subsection{Ideal to Real fluid and gas behaviour : Departure functions }

The NASA polynomials only capture ideal gas entropy, enthalpy and thermal capacity. To capture the deviations of these quantities from ideal gas behavior to obtain the real behaviour, we can use the Peng-Robinson EOS, and compute \emph{departure functions}:
\[
H^{\text{dep}} = H^{\text{real}} - H^{\text{ideal}}, \quad S^{\text{dep}} = S^{\text{real}} - S^{\text{ideal}}.
\]
These corrections are derived analytically from the EOS and depend on \( Z \), \( a \), \( b \), and \( \frac{d(a\alpha)}{dT} \). The formulas are as follows : 

$$H^{dep} = RT (Z - 1) + \frac{T \frac{da(T)}{dT} - a(T)}{2\sqrt{2} b} \ln \left( \frac{Z + (1 + \sqrt{2})B}{Z + (1 - \sqrt{2})B} \right)$$

$$S^{dep} = R \ln (Z - B) + \frac{\frac{da(T)}{dT}}{2\sqrt{2} b} \ln \left( \frac{Z + (1 + \sqrt{2})B}{Z + (1 - \sqrt{2})B} \right)$$

where:
\begin{itemize}
    \item $Z$ is the compressibility factor (solved from the cubic form of PR EOS: $Z^3 - (1 - B)Z^2 + (A - 3B^2 - 2B)Z - (AB - B^2 - B^3) = 0$)
    \item $A = \frac{a(T)P}{R^2 T^2}$
    \item $B = \frac{bP}{RT}$
    \item $\frac{da(T)}{dT}$ is the temperature derivative of $a(T)$, computed as:

$$\frac{da(T)}{dT} = a_c \cdot \frac{-2 \kappa}{T_c^{0.5}} \left[ 1 + \kappa \left(1 - \sqrt{\frac{T}{T_c}}\right) \right] \left(1 - \sqrt{\frac{T}{T_c}}\right)$$
\end{itemize}

In these expressions, the first term accounts for the volume effect, while the second term corrects for the temperature dependence of the attractive forces.

Real-fluid enthalpy and entropy are then expressed as:
\[
H(T, P) = H^{\text{ideal}}(T) + H^{\text{dep}}(T, P, Z, a, b, \frac{d(a\alpha)}{dT}),
\]
\[
S(T, P) = S^{\text{ideal}}(T, P) + S^{\text{dep}}(T, P, Z, a, b, \frac{d(a\alpha)}{dT}).
\]

\subsection{Phase equilibrium and saturation pressure}

Another important element to reconstruct the vapor construction cycle is the relationship between temperature and pressure at a phase equilibrium. At equilibrium, temperature and pressure are not independent variables, as one uniquely determines the other. It is therefore important to be able to get a certain $P_{sat}$ for a given $T$ at saturation, and vice-versa. 

\subsubsection{Chemical Potential and Fugacity}

The \emph{chemical potential} \( \mu \) is defined as the change in Gibbs free energy \( G \) with respect to the number of moles:
\[
\mu = \left( \frac{\partial G}{\partial n} \right)_{T,P}.
\]
It governs mass transfer, phase change, and chemical reaction. At phase equilibrium, the chemical potentials in coexisting phases are equal:
\[
\mu^{\text{liq}} = \mu^{\text{vap}}.
\]

To evaluate chemical potential in real fluids, we introduce the \emph{fugacity} \( f \), related to pressure through the \emph{fugacity coefficient} \( \phi \):
\[
f = \phi P.
\]
For an ideal gas, \( \phi = 1 \), so \( f = P \). For real fluids, \( \phi \neq 1 \), and is computed from the EOS. The chemical potential can then be written as:
\[
\mu(T, P) = \mu^{\text{ideal}}(T, P^\circ) + RT \ln\left( \frac{f}{P^\circ} \right),
\]
where \( P^\circ \) is a reference pressure.

\subsubsection{P and T relationship at phase equilibrium}

At a given temperature, the vapor and liquid phases are in equilibrium when their fugacities match:
\[
f^{\text{liq}}(T, P_{\text{sat}}) = f^{\text{vap}}(T, P_{\text{sat}}),
\]
which implies
\[
\phi^{\text{liq}}(T, P_{\text{sat}}) = \phi^{\text{vap}}(T, P_{\text{sat}}).
\]
Solving this equation yields the saturation pressure \( P_{\text{sat}} \) at temperature \( T \), a key quantity for modeling phase behavior in refrigerant cycles.

The Peng-Robinson EOS gives a relationship between $\phi$ and $Z$, as follows : 

$$ \ln \phi = Z - 1 - \ln(Z - B) - \frac{A}{2\sqrt{2}B} \ln \left( \frac{Z + (1 + \sqrt{2})B}{Z + (1 - \sqrt{2})B} \right)$$

with the $A$ and $B$ parameters being the same ones presented earlier. We can therefore for example get any saturation pressure $P_{sat}$ from $T$ as follows :

\begin{algorithm}[H]
    \caption{Calculation of Saturation Pressure ($P_{sat}$) at a Given Temperature ($T$)}
    \label{alg:psat_calculation}
    \begin{algorithmic}[1]
        \Require Given Temperature $T$, Tolerance $\epsilon$, EOS Parameters $a(T)$ and $b$.
        \Ensure Saturation Pressure $P_{sat}$ at temperature $T$.

        \State Choose an initial trial pressure $P_{trial}$
        \State Calculate EOS Parameters $a(T)$ and $b$ at the given $T$.

        \While{$|\Delta| \geq \epsilon$}
            \State Calculate $A = \frac{a(T) P_{trial}}{(RT)^2}$ and $B = \frac{b P_{trial}}{RT}$
            \State Solve the cubic equation for $Z$: $Z^3 + (B-1)Z^2 + (A-3B^2-2B)Z + (B^3+B^2-AB) = 0$
            \State Identify $Z^L$ (smallest positive root) and $Z^V$ (largest positive root).
            \If{no three real roots exist}
                \State Adjust $P_{trial}$ (e.g., increase if no liquid root, decrease if no vapor root) and \textbf{continue} to next iteration of the While loop. \Comment{Handle cases where $P_{trial}$ is outside the two-phase region}
            \EndIf
            \State Calculate $\ln \phi^L$ using $Z^L$, $A$, $B$.
            \State Calculate $\ln \phi^V$ using $Z^V$, $A$, $B$.
            \State Calculate $\Delta = \ln \phi^L - \ln \phi^V$.

            \If{$|\Delta| < \epsilon$}
                \State $P_{sat} \gets P_{trial}$
                \State \textbf{Break} \Comment{Equilibrium reached}
            \Else
                \State Adjust $P_{trial}$ using a numerical root-finding algorithm (e.g., Secant, Newton-Raphson) targeting $\Delta = 0$.
                \Comment{If $\Delta > 0$, $P_{trial}$ is likely too low; if $\Delta < 0$, $P_{trial}$ is likely too high.}
            \EndIf
        \EndWhile
        \State \Return $P_{sat}$
    \end{algorithmic}
\end{algorithm}

\subsection{Cycle simulation and COP and $Q_{vol}$ calculation}

With methods to calculate $H(T,P)$ and $S(T,P)$ as well as $P_{sat}$ for any $T$ (combining ideal gas contributions from NASA polynomials and real fluid corrections from PR EOS via departure functions), the standard VCC can be simulated. 

We first simulate the saturation dome : 

\begin{algorithm}[H]
    \caption{Calculation of Saturation Dome}
    \label{alg:saturation_dome_updated}
    \begin{algorithmic}[1]
        \Require $T_c$, $P_c$, $\omega$, Fluid Properties $\text{fluid\_props}$ (NASA coefficients, molar mass)
        \State Define a range of temperatures $\mathbf{T}$ from $T_{triple}$ to $T_c$.

        \For{each temperature $T$ in $\mathbf{T}$}
            \State Calculate $P_{sat}^{PR}$ at $T$ using a root-finding method (e.g., Secant).
            
            \State Get EOS parameters $a(T)$, $b$, $da/dT$ at $T$.
            \State Solve PR EOS for compressibility factors $Z^L$ and $Z^V$ at $(T, P_{sat}^{PR}, a(T), b)$.
            \If{$Z^L$ or $Z^V$ are invalid (NaN)}
                \State \textbf{Continue} to next $T$.
            \EndIf
            
            \State Calculate specific volumes $v^L = \frac{Z^L R T}{P_{sat}^{PR}}$ and $v^V = \frac{Z^V R T}{P_{sat}^{PR}}$.
            
            \State Calculate ideal gas enthalpy $h_{ig}(T)$ and entropy $s_{ig}(T, P_{sat}^{PR})$ using NASA polynomials.
            \State Calculate departure functions $h_{dep}^L$, $h_{dep}^V$, $s_{dep}^L$, $s_{dep}^V$.
            \State Calculate enthalpies and entropies with $h(T, P) = h^{\text{ideal}}(T) + h^{\text{dep}}$ and $s(T, P) = s^{\text{ideal}}(T, P) + s^{\text{dep}}$.
        \EndFor

        \State Calculate critical point properties as needed.
        \State \Return All calculated saturation property arrays ($P_{sat}, v^L, v^V, h^L, h^V, s^L, s^V$).
    \end{algorithmic}
\end{algorithm}

We can then finally reconstruct the VCC cycle and calculate the COP and $Q_{vol}$ : 

\begin{algorithm}[H]
    \caption{Calculation of VCC State Points, COP, and $Q_{vol}$ (Temperature-Driven)}
    \label{alg:vcc_calc_updated}
    \begin{algorithmic}[1]
        \Require Evaporator Temperature $T_{evap}$, Condenser Temperature $T_{cond}$, Fluid Properties $\text{params}$, Pre-calculated PR+NASA Saturation Dome Data $\text{dome\_data\_pr}$

        \State Extract and prepare valid temperature-saturation property data ($\mathbf{T}_{sat}$, $\mathbf{P}_{sat}$, $\mathbf{h}_L$, $\mathbf{h}_V$, $\mathbf{s}_L$, $\mathbf{s}_V$, $\mathbf{v}_V$) from $\text{dome\_data\_pr}$.
        \State Create interpolation functions for saturation properties based on temperature.

        \Comment{Determine Saturation Pressures}
        \State $P_{evap} \gets \text{interpolate}(T_{evap})$ from $\mathbf{P}_{sat}$
        \State $P_{cond} \gets \text{interpolate}(T_{cond})$ from $\mathbf{P}_{sat}$

        \Comment{Define Cycle State Points}
        \State \textbf{State 1 (Evaporator Outlet - Saturated Vapor):}
        \Statex \quad $h_1 \gets \text{interpolate}(T_{evap})$ from $\mathbf{h}_V$
        \Statex \quad $s_1 \gets \text{interpolate}(T_{evap})$ from $\mathbf{s}_V$
        \Statex \quad $v_1 \gets \text{interpolate}(T_{evap})$ from $\mathbf{v}_V$ \Comment{Get specific volume}
        \Statex \quad $P_1 \gets P_{evap}$

        \State \textbf{State 3 (Condenser Outlet - Saturated Liquid):}
        \Statex \quad $h_3 \gets \text{interpolate}(T_{cond})$ from $\mathbf{h}_L$
        \Statex \quad $P_3 \gets P_{cond}$

        \State \textbf{State 4 (Expansion Valve Outlet - Isenthalpic):}
        \Statex \quad $h_4 \gets h_3$
        \Statex \quad $P_4 \gets P_{evap}$

        \State \textbf{State 2 (Compressor Outlet - Isentropic):}
        \Statex \quad $s_2 \gets s_1$
        \Statex \quad $P_2 \gets P_{cond}$
        \State Find $T_2$ such that the calculated entropy at $(T_2, P_2)$ equals $s_2$.
        \State Calculate $h_2$ using $(T_2, P_2)$ and the PR+NASA model.

        \Comment{Calculate Performance Metrics}
        \State Compressor Work: $W_c = h_2 - h_1$
        \State Refrigeration Effect: $Q_e = h_1 - h_4$
        \State COP $\gets Q_e / W_c$ (if $W_c > 0$ and $Q_e > 0$)
        \State Volumetric Capacity: $Q_{vol} \gets Q_e / v_1$

        \State \textbf{Return} $(h_1, h_2, h_3, h_4)$, $(P_1, P_2, P_3, P_4)$, COP, $Q_{vol}$.
    \end{algorithmic}
\end{algorithm}

We can then compute
\[
COP = \frac{\text{Refrigerating Effect}}{\text{Compressor Work}} = \frac{h_1 - h_4}{h_2 - h_1} \qquad
    Q_{vol} = \frac{h_1 - h_4}{v_1}
\]

\section{Choice of the property predictor}
\label{appendix:prop_pred_choice}

\subsection{The XGBoost baseline}

Considering that most datasets of properties obtained in this project are on the smaller side, a natural first baseline to test out were tree based models trained on features extracted from the SMILES representation of molecules.

We train an XGBoost model with the following hyperparameters : 
\begin{itemize}
    \item Estimator number : 1000
    \item Learning rate : 0.1
    \item Maximum depth 6
    \item Subsample 0.8
    \item Columns (fraction) sampled by tree : 0.8
    \item Patience 10
    \item Early stopping metric : Mean Absolute Error
\end{itemize}

Using cheminformatics libraries such as RDKit, we extract multiple chemical descriptors from SMILES into a large feature vector : Morgan Fingerprints; Molecular Descriptors such as molecular weight, topological polar surface area (TPSA), LogP, and connectivity indices; Fragment Counts: functional group presence is encoded through SMARTS pattern matching, counting occurrences of chemically relevant substructures (e.g., hydroxyl groups, aromatic rings, halogens).

\subsection{The SMI-TED Model}

We end-to-end finetune the SMI-TED foundation model \citep{Soares2025} on our property datasets, leveraging the learned encoded latent representations, and adding an MLP head, following the architecture and the code from the authors. SMI-TED hyperparameters are the same as for the final property predictors, and is detailed in Appendix \ref{appendix:smi_ted_final_hparams}. 

\subsection{Experiments and results}

We compare the two models on our critical temperature dataset, as well as an additional testing set from the CoolProp database. Details on dataset preprocessing are in the property prediction training details in Appendix \ref{appendix:prop_pred_training_details}.

We compare results with SMILES augmentation, adding 4 additional graph traversals to the initial dataset SMILES, for a total of \textbf{5}, as it ensured that >95\% of all molecules had that many graph traversals and avoided dataset imbalance. 

\begin{table}[h!]
    \centering
    \caption{Percentage of molecules with 5 unique graph traversals}
    \label{tab:property_coverage}
    \begin{tabular}{lc}
        \toprule
        \textbf{Target Property} & \textbf{Percentage of Molecules (\%)} \\
        \midrule
        Tc (Critical Temperature) & 97.8\%    \\
        Pc (Critical Pressure) & 97.9\% \\
        $\omega$ (Acentric Factor) & 97.9\%  \\
        Nasa Polynomials & 99.2\% \\
        Lower Flammability Limit (LFL) & 96.0\% \\
        k(OH) Rate Constant & 96.4\% \\
        Radiative Efficiency (RE) & 99.8\% \\
        \bottomrule
    \end{tabular}
\end{table}

Results are compared and averaged on 5 random seeds.

\begin{table}[h!]
    \centering
    \caption{Mean Prediction Performance of XGBoost vs. SMI-TED}
    \label{tab:mean_performance_2}
    \begin{tabular}{ll ccc ccc}
        \toprule
        \multirow{2}{*}{\textbf{Dataset}} & \multirow{2}{*}{\textbf{Augmentation}} & \multicolumn{3}{c}{\textbf{XGBoost}} & \multicolumn{3}{c}{\textbf{SMI-TED}} \\
        \cmidrule(lr){3-5} \cmidrule(lr){6-8}
        & & R$^2$ & MAE & RMSE & R$^2$ & MAE & RMSE \\
        \midrule
        \multirow{2}{*}{Test Set} 
        & No  & 0.946 & 17.86 & 33.00 & 0.938 & 15.94 & 34.64 \\
        & Yes & 0.942 & 17.58 & 33.32 & \textbf{0.946} & \textbf{14.40} & \textbf{31.06} \\
        \midrule
        \multirow{2}{*}{CoolProp} 
        & No  & 0.554 & 48.92 & 106.80 & \textbf{0.944} &19.72 & 33.18 \\
        & Yes & 0.420 & 56.16 & 121.68 & 0.938 & \textbf{17.94} & \textbf{32.68} \\
        \bottomrule
    \end{tabular}
\end{table}

Not only do the SMILES augmented end-to-end finetuned SMI-TED, avoid the need for handcrafted features, the models showcase better prediction accuracy and especially, better generalization ability on the CoolProp dataset. This shows that despite being trained on very diverse data, the SMI-TED model generalizes better to target refrigerant molecules. This setup is chosen for the final property prediction architecture.

\newpage
\section{Supervised fine-tuning stage}
\label{sft:llama}

\subsection{Data curation}
We consider the concatenation of Pubchem Compounds \cite{kim2021pubchem}, ChEMBL \cite{chembl} and SureChEMBL \cite{papadatos2016surechembl}, amounting to a total of $1.47 \cdot 10^8$ smiles. We apply a filtering pipeline based on the following criteria: (1) excluding molecules with undesired substructures from a predefined list; (2) salts are removed with RDKit's \texttt{SaltRemover} ; (3) in case of fragments, keep only the largest one in the molecule with RDKit's \texttt{LargestFragmentChooser}; (4) remove any atom mapping numbers e.g. \texttt{[CH3:1] → CH3}; (5) neutralize the molecule with RDKit's \texttt{Uncharger}; (6) Normalize the molecule with RDKit's \texttt{Normalizer}, handling aromatization, kekulization, normalization of tautomers, charges, and representations. The output subset from this pipeline counts $3.72 \cdot 10^7$ smiles ($25.3 \%$ of the original dataset, hence $74.7 \%$ rejected structures).

\subsection{Training parameters}

The SFT training employs the following hyperparameters :
\begin{itemize}
    \item Base Model: Llama-3.2-1B
    \item Batch Size: 256 (with gradient accumulation: $\frac{256}{\text{max\_gpu\_batch\_size}}$)
    \item GPU batch size: 16
    \item Hardware: $4 \times$ NVIDIA A100:80GB
    \item Learning Rate: $\alpha = 2 \times 10^{-5}$ with linear scheduling
    \item Optimizer: AdamW with $\beta_1 = 0.9$, $\beta_2 = 0.999$, $\epsilon = 10^{-8}$
    \item Sequence Length: maximum 256 tokens
    \item Training Epochs: 1 epoch over the entire dataset
\end{itemize}

Let $\mathcal{D}_{SFT} = \{\mathbf{x}\}_{i=1}^N$ be the molecular dataset, $\theta$ the model parameters, $\mathcal{T}^{(i)}$ the length of sequence $i$, and $p_\theta(x_t^{(i)} | x_{<t}^{(i)})$ the probability of generating token $x_t^{(i)}$ given the previous tokens $x_{<t}^{(i)} = (x_1^{(i)}, \ldots, x_{t-1}^{(i)})$.

During training, the loss is computed only on the SMILES tokens, not on the property conditioning tokens. This is achieved through attention masking:

$$
    \mathcal{L}_{\text{SFT}} = -\frac{1}{N} \sum_{i=1}^{N} \sum_{t}^{\mathcal{T}_{\text{SMILES}}^{(i)}} \log p_\theta(x_t^{(i)} | x_{<t}^{(i)})
$$
where $\mathcal{T}_{\text{SMILES}}^{(i)}$ represents the set of positions corresponding to SMILES tokens in sequence $i$.

\section{Training property predictors}
\label{appendix:prop_pred_training_details}

\subsection{Dataset processing and splits}

\textbf{Smiles canonicalization} is not performed to avoid loosing data from multiple data entry points which could have relevant information. 

\textbf{Smiles augmentation} is performed, according to results in \ref{appendix:prop_pred_choice} showing its benefits to model generalization. For each molecule in the training set, four additional SMILES traversals are generated. 

\textbf{Dataset cleaning} : \textbf{Metals and metalloids} are discarded if they exist in the data as they are too far out of distribution from our molecules of interest. This is mostly of use for the thermodynamic properties dataset which is a concatenation of multiple existing datasets extracted from the internet and not already curated. \textbf{Single elemental atoms} or ions are also discarded as their properties are often very different from the rest of the molecules and do not represent structures of interest, especially in the case of refrigerant discovery.

\textbf{Scaffolds} Separation of training and testing by molecular scaffolds, also known as core molecular structures, is a common technique to better verify the generalization properties of the model. Molecules with the same scaffolds are grouped together to avoid having too similar structures in different dataset splits. We therefore separate train/validation and test splits by Bemis-Murcko scaffolds, using the RdKit library.

The train / test / validation split is 70/15/15.

Final dataset sizes and splits are as follows : 

\begin{table}[h!]
    \centering
    \caption{Dataset split sizes for different target properties}
    \label{tab:dataset_sizes}
    \begin{tabular}{lcccc}
        \toprule
        \textbf{Dataset sizes} & \textbf{Train} & \textbf{Augmented Train} & \textbf{Validation} & \textbf{Test} \\
        \midrule
        Tc &   8079 & 39615 & 1358 & 1612 \\
        Pc   &  8322     &  40816     &    1329            &   1409 \\
        $\omega$ &  4940     &  24269    & 1074               &  968 \\
        Nasa Polynomials & 11924 & 59435 & 2569 & 2309 \\
        \midrule
        RE & 55938  & 279615 & 11676 & 14380 \\
        k(OH) constant    & 937  & 4552 & 105 & 131  \\
        \midrule
        LFL & 1010 & 4863 & 180 & 171 \\
        \bottomrule
    \end{tabular}
\end{table}

\begin{table}[h!]
    \centering
    \caption{External test sets for model evaluation}
    \label{tab:external_test_sets}
    \begin{tabular}{lcc}
        \toprule
        \textbf{Property} & \textbf{External Dataset} & \textbf{Test Size} \\
        \midrule
        Tc & CoolProp & 106 \\
        Pc & CoolProp & 107 \\
        $\omega$ & CoolProp & 107 \\
        GWP & IPCC Reports & 220 \\
        \bottomrule
    \end{tabular}
\end{table}

\subsection{SMI-TED hyperparameter details}
\label{appendix:smi_ted_final_hparams}

Hyperparameter details : 

\begin{itemize}
    \item Batch size: 256
    \item Learning rate: $3 \times 10^{-5}$
    \item Learning rate multiplier: 1.0
    \item Optimizer: AdamW with $\beta_1 = 0.9$, $\beta_2 = 0.99$
    \item Loss function: Mean Absolute Error (MAE)
    \item Early stopping epochs : 10
\end{itemize}

\subsection{Group contribution for kOH constant calculation}
\label{atkinson_gc_details}

Group contribution methods estimate molecular properties by decomposing a molecule into predefined substructures (or “groups”) and summing their individual contributions. Each group has an associated rate or factor derived from experimental data, and the total property (e.g., reaction rate) is the sum of the contributions from all relevant groups in the molecule. This enables rapid predictions without full quantum chemical calculations.

We implemented a group contribution method based \cite{Kwok1995Estimation} and the EPA’s AOPWIN software refinements \citep{US_EPA_EPI_2012}, which estimates the OH radical reaction rate constants by summing contributions from four main pathways:

\begin{itemize}
    \item H-abstraction from C--H, O--H, or S--H bonds
    \item OH addition to unsaturated bonds (\chemfig{C=C}, \chemfig{C~C})
    \item OH addition to aromatic rings
    \item OH interactions with specific functional groups
\end{itemize}

The total rate constant is modeled as:

\begin{equation*}
    k_{{OH,tot}} = k_{\text{H abs}} + k_{\text{OH add, unsat}} + k_{\text{OH add, ring}} + k_{\text{OH func. group}}
\end{equation*}

Each term is computed using empirically fitted base rate constants and multiplicative modifiers called \textbf{substituent factors}. For example, in H-abstraction from saturated carbon atoms, the rate is determined using formulas like:

\begin{equation*}
    k({X-CH2-Y}) = k_{\text{sec}} \cdot F(X) \cdot F(Y), \quad
    k({X-CH<Y,Z}) = k_{\text{tert}} \cdot F(X) \cdot F(Y) \cdot F(Z)
\end{equation*}

Here, $k_{\text{sec}}$ and $k_{\text{tert}}$ are base rate constants for secondary and tertiary hydrogens, and $F(\cdot)$ are substituent factors that encode how reactive neighboring groups are. 

\section{Reinforcement Learning Fine-Tuning stage}
\label{appendix:grpo_details}

\subsection{Reward functions}
\label{appendix:reward_functions}

\subsubsection{Molecule Validity}

As the base model has been finetuned to generate valid SMILES, we need to ensure that our RL finetuning does not make the distribution diverge to the point it stops generating valid molecules. We therefore build a reward for molecule validity alongside property-specific rewards:
$$R_{\text{final}}(\mathbf{x}) = R_{\text{validity}}(\mathbf{x}) + R_{\text{properties}}(\mathbf{x})$$
where:
$$R_{\text{validity}}(\mathbf{x}) = \begin{cases} 
1.0 & \text{if SMILES is valid and parseable} \\
0.0 & \text{otherwise}
\end{cases}$$
$$R_{\text{properties}}(\mathbf{x}) = \begin{cases} 
\text{property-based score} & \text{if SMILES is valid and parseable} \\
0.0 & \text{otherwise}
\end{cases}$$

This additive formulation ensures that invalid molecules receive zero total reward (since both validity and property rewards are zero), while valid molecules receive their property-based score plus a +1.0 validity bonus, strongly encouraging the model to maintain chemical validity while optimizing for desired characteristics.

We verify two elements for validity :
\begin{itemize}
    \item We ensure the SMILES is a valid molecule through the RdKit library
    \item We also ensure that each atom in the the molecule has the correct valence (also through RdKit). We consider molecules with radicals to be invalid as they are very unstable. 
\end{itemize}

This second element in the validity reward was added after later analysis that, as the model strayed from the initial distribution, it tended to generate molecules with radicals (i.e with squared brackets in the molecule), which had good properties, but are very unstable in practice. Stability, despite not being its own reward or a property analyzed in detail in this project, is key for refrigerants. 

\subsubsection{Critical temperature}

Critical temperature is an essential property of a refrigerant and is used here, as in \citet{Kazakov2012ComputationalDO} and \citet{McLinden2017}, as the main thermodynamic criterion for the selection of molecules. They adopt lower bounds of 300K for the minimal operating temperature of the condenser in the cycle, as well as a generous upper bound of 550K, explaining that machines with centrifugal compressors can work with refrigerants with $T_c$ up to 470K+. The ideal range however, has been shown, and is explicitly given in \citet{McLinden2017} as [320K - 420K].

We therefore build a tiered exponentially decreasing reward as such, with a reward of 1 in the ideal range, and attaining 0.5 in the larger 300-550 range. 

Let $T$ be the temperature in Kelvin. The function $f(T)$ is defined as follows:

\[
f(T) = \begin{cases}
    e^{-k_{\text{left}} (T_{\text{plateau\_min}} - T)} & \text{if } T < T_{\text{plateau\_min}} \\
    1.0 & \text{if } T_{\text{plateau\_min}} \le T \le T_{\text{plateau\_max}} \\
    e^{-k_{\text{right}} (T - T_{\text{plateau\_max}})} & \text{if } T_{\text{plateau\_max}} < T \le T_{\text{target\_max\_temp}} \\
    R_{\text{target}} \cdot e^{-k_{\text{left}} (T - T_{\text{target\_max\_temp}})} & \text{if } T > T_{\text{target\_max\_temp}}
\end{cases}
\]

where $T_{\text{plateau\_min} = 320K}$, $T_{\text{plateau\_max} = 420K}$, $T_{\text{target\_min\_temp}} = 300K$, $T_{\text{target\_max\_temp}} = 550K $, the  target reward at boundary temperature $R_{\text{target}} = 0.5$ and $k_{\text{left}} = \dfrac{-\ln(R_{\text{target}})}{T_{\text{plateau\_min}} - T_{\text{target\_min\_temp}}}$ and  $k_{\text{right}} = \dfrac{-\ln(R_{\text{target}})}{T_{\text{target\_max\_temp}} - T_{\text{plateau\_max}}}$

\subsubsection{Molecular length}

The length penalty is essential to obtain tractable molecules that are not completely out of distribution. Moreover, refrigerants are small molecules. Empirical results and thermodynamic results show that refrigerants have rarely over 15-18 atoms. In their work, \citet{Kazakov2012ComputationalDO} filter PubChem for under 15 atoms while in their final work \citep{McLinden2017}, a larger range of up to 18 atoms are analyzed. 

We use RdKit to count the number of atoms. A custom gaussian-like reward function with a plateau and custom curve steepness is used, to avoid generating individual atoms, as well as molecules over 18 atoms. 

Let $x$ be the measured property value (e.g., number of atoms). The function $g(x)$ is defined as follows:

\[
g(x) = \begin{cases}
    e^{-0.5 \left( \frac{|x - P_{\text{start}}|}{\sigma} \right)^{S}} & \text{if } x < P_{\text{start}} \\
    1.0 & \text{if } P_{\text{start}} \le x \le P_{\text{end}} \\
    e^{-0.5 \left( \frac{|x - P_{\text{end}}|}{\sigma} \right)^{S}} & \text{if } x > P_{\text{end}}
\end{cases}
\]

where  the start of the ideal range $P_{\text{start}} = 7$,the end of the ideal range $P_{\text{end}} = 18$, $\sigma = 3$ and $S = 4$.

\subsubsection{COP \& $\mathbf{Q_{vol}}$}

The Coefficient of Performance of the thermodynamic cycle associated to the specific refrigerant molecule is defined at specific evaporator and condenser temperatures. We use the same operating conditions used for AC systems suggested by \citet{McLinden2017} :
\begin{itemize}
    \item $T_{\text{evap}} = 10\,^{\circ}\mathrm{C}$
    \item $T_{\text{cond}} = 40\,^{\circ}\mathrm{C}$
\end{itemize}

Empirically, most generated molecules have a high COP. We therefore build a \textbf{increasing logistic function} around the $Q_{vol}$ such that the midpoint is about a third of the value of R-410, as done in \citet{McLinden2017} for their filtering. The function is as follows : 
\[
R(x) = \frac{L}{1 + e^{-k(x - x_0)}}
\]
where : $L = 1$ is the upper asymptote, $k = 2$ is the steepness and $x_0 = 2$ is the midpoint. An additional constraint is adding on COP, with the reward being 0 if $COP < 5$.

\subsubsection{GWP}

Modern day constraints on refrigerants are still somewhat lenient despite being increasingly restrictive. In their detailed screening paper, \cite{Kazakov2012ComputationalDO}, use a limit of 200, according to recent european guidelines of refrigerants in the automotive industry in Europe. An ideal objective however, is the use of refrigerants with GWP < 1, i.e with lower global warming potentials than CO2. This is the case of HFO's which are the current new viable candidates that are emerging in the industry. 

We therefore want to push our generation towards GWP < 1, we use a decreasing exponential with a plateau. The reward function for GWP, is given by:
$$
R(\text{GWP}) = \begin{cases}
    1 & \text{if } GWP \le \text{plateau\_end} \\
    e^{-\text{decay\_rate} \cdot (GWP - \text{plateau\_end})} & \text{if } GWP > \text{plateau\_end}
\end{cases}
$$

With $\text{plateau\_end} = 0.5$ and $\text{decay\_rate} = 0.15$ This gives a slow decay approaching 0 at 25-30.

\subsubsection{Flammability}

Flammability is also an important factor for refrigerants. We follow, as \cite{Kazakov2012ComputationalDO} and \cite{McLinden2017}, the ASHRAE standard, with molecules with a Lower Flammability Limit of > 0.1 $kg/m^3$ considered in a good range and almost non flammable. 

We use the same increasing logistic function as for $Q_{vol}$ pressure with $L = 1.0$, $ x_0 = 0.1 $ and $ k = 70 $

\subsubsection{The diversity reward}

The diversity reward aims at penalizing molecules which have appeared often in the recent batches. This seems crucial after initial testing showed lack of exploration and mode collapse at training. 

A memory size limit is chosen depending on how far back one wants to penalize generation. The diversity reward stores canonical SMILES and their counts, timestamps, and fingerprints, which are stored in a FIFO manner. The diversity reward works through two different mechanisms:
\begin{itemize}
    \item A molecule count penalty based on how many times the SMILES has appeared in the memory dictionary.
    \item A Tanimoto similarity penalty based on fingerprints.
\end{itemize}

\paragraph{Molecule Count Penalty (Exponential)}
The core of the penalty system is an exponential function that increases the penalty as the count of a molecule rises. The penalty for a molecule that has appeared $c$ times is calculated with the formula:
$$
P(c) = 1 - e^{-k \cdot (c-1)}
$$
Where $c$ is the count of how many times the exact same molecule has been seen and $k$ is a rate constant, which is set to 0.5.

\paragraph{Tanimoto Similarity Penalty}
When a new molecule is not an exact match but is similar to a molecule already in memory, the penalty is scaled by the Tanimoto similarity score. This ensures that very similar molecules are penalized almost as much as exact matches. The formula for the penalty due to a similar molecule is:
$$
P_{\text{similar}} = P(c_{\text{similar}}) \cdot S_{\text{Tanimoto}}
$$
Where $P(c_{\text{similar}})$ is the penalty calculated above for the similar molecule that is already in memory, which has been seen $c_{\text{similar}}$ times and $S_{\text{Tanimoto}}$ is the Tanimoto similarity score (a value between 0.0 and 1.0) between the new molecule and the similar molecule from memory.

\paragraph{Final Diversity Reward}
The system calculates a penalty for the new molecule being an exact match ($P_{\text{exact}}$) and for its similarity to all molecules in memory that are above the similarity threshold ($P_{\text{similar},1}, P_{\text{similar},2}, \dots$). To determine the final penalty, it takes the single harshest (maximum) penalty from all possibilities:
$$
P_{\text{max}} = \max(P_{\text{exact}}, P_{\text{similar},1}, P_{\text{similar},2}, \dots)
$$
The final diversity reward ($R_{\text{diversity}}$) is then calculated as 1 minus this maximum penalty. The reward is a value between 0.0 (for a highly unoriginal molecule) and 1.0 (for a completely novel molecule).
$$
R_{\text{diversity}} = 1 - P_{\text{max}}
$$

\subsection{GRPO training parameters}

\begin{itemize}
    \item Learning rate : $5e^{-7}$ with linear scheduler
    \item KL $\beta$ : 0
    \item Batch size : 16
    \item Number of generations per batch : 4
    \item Optimizer : AdamW with $\beta_1 = 0.9$, $\beta_2 = 0.999$
    \item Patience 500 with minimum $\delta = 0.1$
    \item Temperature = 1
    \item Top p sampling : 0.9
\end{itemize}

\textbf{Reward weights} :
\begin{table}[h!]
    \centering
    \caption{Reward weights used in training}
    \label{tab:reward_weights}
    \begin{tabular}{lc}
        \toprule
        \textbf{Metric} & \textbf{Weight} \\
        \midrule
        COP \& $Q_{vol}$ & 4.0 \\
        $T_c$ & 4.0 \\
        Molecular length & 1.0 \\
        GWP & 0.5 \\
        LFL & 0.5 \\
        \bottomrule
    \end{tabular}
\end{table}

The choice of the weights was made through simple parameter search and tuning, and reflects the difficulty of achieving the different desired properties. $Q_{vol}$ and $T_c$ are harder to optimize for but essential properties and therefore are given higher weights.

\subsection{GRPO Inference details}

Molecule generation at inference time is generated through 'Top-p' nucleus sampling with a threshold of 0.9, as well as a temperature of 1. 

\section{Property prediction results}
\label{appendix:prop_pred_results}

\subsection{Coefficient of Performance (COP)}

\begin{table}[h!]
    \centering
    \caption{Prediction Performance of Thermodynamic Properties (Test Set)}
    \label{tab:summary_performance_updated}
    \begin{tabular}{l ccc}
        \toprule
        \textbf{Setup} & \textbf{R$^2$} & \textbf{MAE} & \textbf{RMSE} \\
        \midrule
        Tc                    & 0.975 & 12.4696 & 21.835 \\
        Pc                    & 0.876 & 148931 & 568426 \\
        $\omega$              & 0.8749 & 0.0398 & 0.0868 \\
        \midrule
        \textbf{Overall Nasa Polynomial Mean} & \textbf{0.9647} & \textbf{237.2239} & \textbf{856.0378} \\
        \addlinespace
        \quad Target: a1      & 0.9576 & 0.1223 & 0.1746 \\
        \quad Target: a2      & 0.9904 & 0.00072 & 0.00099 \\
        \quad Target: a3      & 0.9864 & $7.42 \times 10^{-7}$ & $1.11 \times 10^{-6}$ \\
        \quad Target: a4      & 0.9829 & $3.37 \times 10^{-10}$ & $5.06 \times 10^{-10}$ \\
        \quad Target: a5      & 0.9800 & $5.22 \times 10^{-14}$ & $7.82 \times 10^{-14}$ \\
        \quad Target: a6      & 0.9971 & 1659.9358 & 2264.8633 \\
        \quad Target: a7      & 0.8588 & 0.5101 & 0.7571 \\
        \bottomrule
    \end{tabular}
\end{table}

To test nasa polynomials and the Peng-Robinson model, we measure errors on the saturation dome construction and the associated COP prediction for the \textbf{CoolProp} \citep{CoolProp2025} fluids. The COP is calculated for condenser and evaporator temperatures of 10°C and 40°C. 

\begin{table}[h!]
    \centering
    \caption{Thermodynamic Property Prediction on CoolProp}
    \label{tab:coolprop_properties}
    \begin{tabular}{lccc}
        \toprule
        \textbf{Property} & \textbf{R$^2$} & \textbf{MAE} & \textbf{RMSE} \\
        \midrule
        Tc & 0.926 & 20.69 & 37.60 \\
        Pc & 0.769 & 502134 & 1451979 \\
        $\omega$ & 0.636 & 0.057 & 0.125 \\
        \bottomrule
    \end{tabular}
\end{table}

\begin{table}[h!]
    \centering
    \caption{Saturation Dome RMSE and COP MAE on CoolProp}
    \label{tab:coolprop_system_metrics}
    \begin{tabular}{lccccc}
        \toprule
        \textbf{Metric} & \textbf{$N_{samples}$} & \textbf{Mean} & \textbf{Median} & \textbf{Std Dev} \\
        \midrule
        Dome latent heat RMSE ($J$) & 110 & 30.9 & 17.5 & 41.4 \\
        $|\Delta \mathrm{COP}|$ $(\varnothing)$ & 82 & 0.252 & 0.12 & 0.343 \\
        \bottomrule
    \end{tabular}
\end{table}

\subsection{Global Warming Potential (GWP)}

\subsubsection{Radiative Efficiency (RE)}

The SMI-TED finetuned model exhibits good performance on radiative efficiency prediction, as for the other properties. 

\begin{table}[h!]
    \centering
    \caption{Prediction Performance of radiative efficiency}
    \label{tab:summary_performance}
        \begin{tabular}{l ccc}
        \toprule
        \textbf{Setup} & \multicolumn{3}{c}{\textbf{Test Set}} \\
        \cmidrule(lr){2-4}
                                          & \textbf{R$^2$} & \textbf{MAE} & \textbf{RMSE}\\
        \midrule
        RE & 0.91 & 0.016 & 0.025 \\
        \bottomrule
    \end{tabular}
\end{table}

\subsubsection{k(OH) constant}

We compare three methods for k(OH) prediction before testing on the database of 220 GWP values :
\begin{enumerate}
    \item Direct SMI-TED prediction of k(OH) through the \citet{McGillen2020DatabaseFT} dataset
    \item Using a custom group contribution (GC) method inspired by the \citet{Kwok1995Estimation} paper and notes from the \citet{US_EPA_EPI_2012} AOPWIN software.
    \item Training a SMI-TED model on correction factors between the ground truth dataset and the group contribution method
\end{enumerate}

\textbf{Important remark} : the \citep{Kwok1995Estimation} group contribution method gives a reaction rate of 0 for molecules not reacting according to the 4 main reaction pathways detailed in the method. This usually means the true reaction rate is very slow. We decide to put these values at the arbitrarily small $ k_{OH} = 10^{-20}$. For the SMI-TED models, considering the very similar results, training was done on 5 random seeds.

\textbf{Generalization ability}: For the k(OH) constant, more than just good predictions on the \citet{McGillen2020DatabaseFT} dataset, we try to consider generalization ability. The group contribution method we implemented is adapted from the AOPWIN software from the EPA (which is not open source) \citep{US_EPA_EPI_2012}, which has been tuned on many different molecules and is considered to have  good generalization ability.  We therefore compare our own group contribution implementation based on open source AOPWIN notes, the SMI-TED k(OH) prediction model the error correction model against the AOPWIN group contribution predicted values (as a form of \textit{'ground truth'}) on ~72k molecules extracted from PubChem following the methodology in \citet{Kazakov2012ComputationalDO}:  Molecules with less than 18 atoms and only C, F, H, S, Cl, Br atoms.

We analyze 3 elements : (1) The quality of our group contribution reconstruction (2) the generalization ability of the SMI-TED k(OH) prediction model (3) How much the error correction model ends up deviating from the AOPWIN method.

Results are averaged on 5 random seeds for the SMI-TED based models. 

\begin{table}[h!]
    \centering
        \caption{Average k(OH) prediction performance comparison with AOPWIN}
    \label{tab:summary_kOH_performance}
    \resizebox{\textwidth}{!}{
    \begin{tabular}{lcccccccc}
        \toprule
        \multicolumn{1}{c}{} & \multicolumn{5}{c}{Log scale metrics} & \multicolumn{3}{c}{Original scale metrics} \\
        \midrule
        \textbf{Setup} & \textbf{R$^2$} & \textbf{$R_{pearson}$} & \textbf{$\tau_{kendall}$} & \textbf{MALE} & \textbf{10$^{\text{MALE}}$} & \textbf{\% <2x} & \textbf{<5x} & \textbf{<10x} \\
        \midrule
        Our GC method & \textbf{0.6623} & \textbf{0.8138} & \textbf{0.6338} & \textbf{0.4094} & \textbf{2.5668} & \textbf{61.28} & \textbf{77.92} & \textbf{87.07} \\
        SMI-TED k(OH) Pred. & 0.4301 & 0.6426 & 0.5004 & 0.6697 & 4.7074 & 30.64 & 61.55 & 77.03  \\
        Our GC + kOH Correction & 0.5849 & 0.7643 & 0.5569 & 0.6052 & 4.0641 & 38.67 & 67.56 & 79.85\\ 
        \bottomrule
    \end{tabular}
    }
\end{table}

\subsubsection{Total GWP prediction}

\begin{table}[h!]
    \centering
    \caption{GWP prediction performance comparison}
    \label{tab:summary_GWP_performance}
    \resizebox{\textwidth}{!}{
    \begin{tabular}{lcccccc}
        \toprule
        \textbf{Setup} & \textbf{R$^2$} & \textbf{Factor error} ($10^{\text{RMSE}_{\log_{10}}}$) & \textbf{\% <2x} & \textbf{\% <5x} & \textbf{\% <10x} \\
        \midrule
        Our GC method & \textbf{0.535} & \textbf{11.46} & \textbf{37.73} & \textbf{65.45} & \textbf{75.91} \\
        SMI-TED k(OH) Pred. & 0.079 & 12.18 & 24.16 & 49.89 & 66.64 \\
        Our GC + kOH Correction & 0.434 & 12.03 & 34.17 & 59.84 & 73.84 \\
        \bottomrule
    \end{tabular}
    }
\end{table}

\paragraph{Conclusion} It seems as if the implemented group contribution method, based on the AOPWIN ('ground truth baseline') software achieves >60\% of predictions with an error under a factor 2 of the software, showing that, despite some edge reaction cases not detailed in their work, we managed to rebuild the AOPWIN method pretty well. The group contribution's generalization capabilities seem to enable it have better performance in GWP prediction compared to the direct SMI-TED k(OH) predictors. In fact, the error correction term actually impedes generalization and does not help with prediction. 

\textbf{We therefore deem it best to use the group contribution by itself for k(OH) constant prediction.}

\subsection{Flammability}

\begin{table}[h!]
    \centering
    \caption{Prediction Performance Flammability }
    \label{tab:summary_performance}
        \begin{tabular}{l ccc}
        \toprule
        \textbf{Setup} & \multicolumn{3}{c}{\textbf{Test Set}} \\
        \cmidrule(lr){2-4}
                                          & \textbf{R$^2$} & \textbf{MAE} & \textbf{RMSE}\\
        \midrule
        LFL & 0.83  & 0.16 & 0.61 \\
        \bottomrule
    \end{tabular}
\end{table}

\section{Final best molecules}
\label{appendix:final_molgen_results}

Only non-PFAS molecules are kept. We add the $Q_{vol} > \frac{Q_{vol-R-410A}}{3}$ filter suggested by \citet{McLinden2017} to ensure reasonable system size, not added before because of the limited amount of molecules satisfying this constraint. A list of all best candidates is given. 

\begin{table}[h!]
\centering
\caption{Physicochemical and Environmental Properties of Best Generated Compounds}
\label{tab:compounds}
\begin{tabular}{lrrrrrrr}
\toprule
\textbf{SMILES} & \textbf{COP} & \textbf{$T_c$ (K)} & \textbf{$p_{evap}$ (kPa)} & \textbf{GWP100} & \textbf{LFL (kg/m$^3$)} & \textbf{$Q_{vol}$} & \textbf{Num Atoms} \\
\midrule
\texttt{C(F)N(F)N(F)F} & 6.71 & 337.82 & 1068.09 & 3.95 & 0.52 & 5.34 & 7 \\
\texttt{C1(F)N(F)N1F} & 7.40 & 350.56 & 932.38 & 0.38 & 0.42 & 5.29 & 6 \\
\texttt{N(F)C(F)N(F)F} & 6.72 & 338.05 & 1050.42 & 0.25 & 0.54 & 5.26 & 7 \\
\texttt{C(F)(N(F)F)N(F)F} & 5.76 & 327.85 & 1182.34 & 0.22 & 0.57 & 4.74 & 8 \\
\texttt{N1(F)N(F)N1F} & 7.72 & 362.07 & 753.04 & 0.42 & 0.44 & 4.54 & 6 \\
\texttt{N1(F)C(F)N1F} & 7.71 & 361.48 & 725.39 & 0.49 & 0.41 & 4.45 & 6 \\
\texttt{C1(F)N(F)O1} & 7.99 & 372.12 & 640.08 & 0.52 & 0.33 & 4.31 & 5 \\
\texttt{NC(F)N(F)F} & 7.81 & 363.11 & 579.40 & 0.27 & 0.43 & 3.99 & 6 \\
\texttt{N1(F)C(F)(N(F)F)O1} & 7.36 & 355.25 & 617.32 & 0.75 & 0.59 & 3.60 & 8 \\
\texttt{N(F)OC(F)N(F)F} & 7.55 & 359.12 & 527.72 & 0.73 & 0.62 & 3.50 & 8 \\
\texttt{C(F)(F)C1(F)N(F)O1} & 7.35 & 356.30 & 593.85 & 0.64 & 0.57 & 3.48 & 8 \\
\texttt{C(F)(C1(F)N(F)O1)F} & 7.44 & 359.69 & 553.55 & 0.64 & 0.56 & 3.31 & 8 \\
\texttt{C1C(F)(N(F)F)N1F} & 7.64 & 366.07 & 511.37 & 0.52 & 0.44 & 3.17 & 8 \\
\texttt{N1C(F)(N(F)F)N1F} & 7.75 & 368.05 & 494.60 & 0.57 & 0.61 & 3.16 & 8 \\
\texttt{C1=C(F)O1} & 8.39 & 403.27 & 393.13 & 0.08 & 0.14 & 3.06 & 4 \\
\texttt{C1(F)N(F)N(F)N1F} & 7.87 & 373.59 & 463.07 & 0.13 & 0.59 & 3.04 & 8 \\
\texttt{C1(F)N(F)N(F)O1} & 8.01 & 376.75 & 424.91 & 0.22 & 0.50 & 2.98 & 7 \\
\texttt{C1(F)C(F)(N(F)F)N1F} & 7.48 & 364.03 & 492.50 & 0.56 & 0.63 & 2.98 & 9 \\
\texttt{N1(F)C(F)N(F)N1F} & 7.88 & 374.77 & 445.63 & 0.16 & 0.60 & 2.95 & 8 \\
\texttt{N1(F)N(F)C(F)N1F} & 7.88 & 375.36 & 438.01 & 0.15 & 0.60 & 2.91 & 8 \\
\texttt{N1(F)C(F)N(F)O1} & 8.02 & 377.78 & 406.60 & 0.23 & 0.51 & 2.88 & 7 \\
\texttt{N1(F)C(F)(N(F)F)N1} & 7.79 & 370.94 & 431.26 & 0.73 & 0.60 & 2.83 & 8 \\
\texttt{N(F)C1(F)N(F)N1F} & 7.91 & 375.60 & 417.94 & 0.53 & 0.62 & 2.83 & 8 \\
\texttt{N1C(F)(N(F)F)O1} & 8.01 & 376.65 & 393.64 & 0.71 & 0.51 & 2.80 & 7 \\
\texttt{N1(F)C(F)(N(F)F)N1F} & 7.65 & 369.09 & 437.30 & 0.58 & 0.68 & 2.78 & 9 \\
\texttt{N(F)C(F)N(F)N(F)F} & 7.69 & 368.44 & 409.19 & 0.23 & 0.68 & 2.76 & 9 \\
\texttt{N1(F)N(F)N(F)N1F} & 7.98 & 380.43 & 406.94 & 0.17 & 0.59 & 2.74 & 8 \\
\texttt{C(F)(C1(F)N(F)N1F)F} & 7.62 & 368.95 & 430.36 & 0.54 & 0.66 & 2.72 & 9 \\
\texttt{C(F)(F)C1(F)N(F)N1F} & 7.62 & 369.56 & 428.57 & 0.54 & 0.66 & 2.72 & 9 \\
\texttt{C(F)C1(F)N(F)N1F} & 7.86 & 377.53 & 401.27 & 0.50 & 0.47 & 2.68 & 8 \\
\texttt{N1(F)N(F)C(F)O1} & 8.10 & 382.82 & 347.83 & 0.25 & 0.50 & 2.57 & 7 \\
\texttt{N1(F)N(F)N1C(F)F} & 7.97 & 381.52 & 373.18 & 0.54 & 0.61 & 2.55 & 8 \\
\texttt{N(F)N(F)C(F)N(F)F} & 7.78 & 373.30 & 365.08 & 0.22 & 0.67 & 2.54 & 9 \\
\texttt{N1(F)C(F)N1N(F)F} & 7.98 & 381.33 & 372.24 & 0.50 & 0.61 & 2.54 & 8 \\
\texttt{C(F)N(F)N(F)N(F)F} & 7.83 & 375.13 & 346.73 & 3.87 & 0.67 & 2.46 & 9 \\
\texttt{N1(F)N(F)N(F)O1} & 8.17 & 389.04 & 334.60 & 0.26 & 0.51 & 2.44 & 7 \\
\texttt{NC(F)N(F)N(F)F} & 8.04 & 379.35 & 311.08 & 0.23 & 0.62 & 2.44 & 8 \\
\texttt{C(F)C1(F)N(F)N(F)1} & 7.88 & 379.72 & 357.74 & 0.57 & 0.49 & 2.43 & 8 \\
\texttt{N1(F)N(C(F)F)N1F} & 8.04 & 384.80 & 343.60 & 0.52 & 0.61 & 2.42 & 8 \\
\texttt{N(F)NC(F)N(F)F} & 8.10 & 382.07 & 301.84 & 0.25 & 0.63 & 2.42 & 8 \\
\texttt{C1(F)N(F)N1N(F)F} & 8.05 & 384.84 & 344.78 & 0.44 & 0.60 & 2.39 & 8 \\
\texttt{N1(F)N(F)C1(F)N(F)F} & 7.79 & 377.65 & 357.47 & 0.52 & 0.67 & 2.37 & 9 \\
\texttt{N(F)C1(F)N(F)O1} & 8.19 & 388.06 & 298.65 & 0.68 & 0.52 & 2.28 & 7 \\
\texttt{C(F)(N(F)F)N(F)N(F)F} & 7.65 & 374.01 & 338.49 & 0.21 & 0.70 & 2.27 & 10 \\
\texttt{C1(F)N(F)C1(F)N(F)F} & 7.79 & 379.55 & 335.69 & 0.53 & 0.64 & 2.24 & 9 \\
\texttt{N1(F)C(F)C1(F)N(F)F} & 7.76 & 379.54 & 338.57 & 0.56 & 0.64 & 2.24 & 9 \\
\texttt{C(F)(F)NC(F)N(F)F} & 7.79 & 377.70 & 306.51 & 0.23 & 0.67 & 2.24 & 9 \\
\texttt{N1(F)N(F)N1N(F)F} & 8.09 & 389.17 & 321.74 & 0.46 & 0.62 & 2.24 & 8 \\
\texttt{N1(C(F)F)N(F)N1F} & 8.11 & 389.52 & 310.28 & 0.54 & 0.60 & 2.24 & 8 \\
\texttt{C1(F)N(F)N1C(F)F} & 8.05 & 388.19 & 309.18 & 0.50 & 0.44 & 2.20 & 8 \\
\texttt{C(F)C(F)N(F)F} & 8.07 & 389.43 & 286.64 & 3.54 & 0.40 & 2.16 & 7 \\
\texttt{CC(F)N(F)F} & 8.16 & 390.68 & 274.20 & 1.88 & 0.23 & 2.13 & 6 \\
\texttt{C(F)(F)OC(F)N(F)F} & 7.71 & 377.07 & 295.55 & 0.69 & 0.67 & 2.12 & 9 \\
\texttt{C(F)(N(F)N(F)F)N(F)F} & 7.73 & 377.42 & 300.02 & 0.21 & 0.70 & 2.08 & 10 \\
\texttt{C(F)(N(F)F)C(F)N(F)F} & 7.59 & 377.37 & 305.02 & 5.07 & 0.73 & 2.07 & 10 \\
\texttt{C(F)(F)OC1(F)N(F)O1} & 7.84 & 378.21 & 283.25 & 0.71 & 0.66 & 2.07 & 9 \\
\texttt{C1=C(F)N1F} & 8.49 & 416.10 & 253.52 & 0.14 & 0.19 & 2.07 & 5 \\
\texttt{N1(F)OC(F)N1F} & 8.31 & 399.74 & 261.63 & 0.20 & 0.50 & 2.05 & 7 \\
\texttt{N1(F)OC(F)(N(F)F)O1} & 7.93 & 382.48 & 275.23 & 0.32 & 0.68 & 2.02 & 9 \\
\texttt{C1(F)OC(F)(N(F)F)N1F} & 7.80 & 380.46 & 282.67 & 0.22 & 0.73 & 2.00 & 10 \\
\texttt{C(F)(C(F)N(F)F)N(F)F} & 7.66 & 379.58 & 286.00 & 4.87 & 0.73 & 1.99 & 10 \\
\bottomrule
\end{tabular}
\end{table}

\end{document}